\documentclass[conference]{IEEEtran}
\usepackage{amsmath,amssymb,amsfonts}
\usepackage{algorithmic}
\usepackage{graphicx}
\usepackage{textcomp}
\usepackage[svgnames]{xcolor}
\usepackage{tikz,tabularx}
\usepackage[hyphens]{url}
\usepackage{fp}
\usepackage{xfp} 

\usepackage[T1]{fontenc}
\usepackage{aecompl}

\usepackage[noadjust,nocompress]{cite}

\usepackage[T1]{fontenc}
\usepackage{caption,subcaption,yfonts,multirow,xspace,enumitem,makecell,pifont,comment}
\usepackage{balance}

\usepackage[bookmarks=true,breaklinks=true,letterpaper=true,colorlinks,citecolor=blue,linkcolor=blue,urlcolor=black]{hyperref}

\newcommand{\Attend}{{Hercules}\xspace}

\newcommand*\mininumbercircled[1]{\tikz[baseline=(char.base)]{
		\node[fill=darkgray, text=white, shape=circle,draw,inner sep=0.1pt] (char) {#1};}}

\def\BibTeX{{\rm B\kern-.05em{\sc i\kern-.025em b}\kern-.08em
    T\kern-.1667em\lower.7ex\hbox{E}\kern-.125emX}}

\title{Enabling Atomic Durability for Persistent Memory with Transiently Persistent CPU Cache} 
\author{\IEEEauthorblockN{Chongnan Ye, Meng Chen, Qisheng Jiang, and Chundong Wang\IEEEauthorrefmark{2}}
	$^\dagger${\em Corresponding author (\href{mailto:cd_wang@outlook.com}{cd\_wang@outlook.com})}
\IEEEauthorblockA{ShanghaiTech University, Shanghai, China}
}

\begin{document}
\maketitle
\thispagestyle{plain}
\pagestyle{plain}


\begin{abstract}
	Persistent memory (pmem) products bring the persistence domain up to the memory level. Intel recently 
	introduced the eADR feature that guarantees to flush data buffered in CPU cache to pmem on a power outage, thereby making the CPU cache a {\em transient persistence domain}. Researchers  
	 have explored how to enable the {\em atomic durability} for applications' in-pmem data. In this paper, we exploit the eADR-supported CPU cache to do so. A modified cache line, until written back to pmem, is a natural redo log copy of the in-pmem data. However, a write-back due to cache replacement or eADR on a crash overwrites the original copy. We accordingly develop \Attend, a hardware logging design for the transaction-level atomic durability, with supportive components installed in CPU cache, memory controller (MC), and pmem. When a transaction commits, \Attend \textit{commits on-chip} its data staying in cache lines. For cache lines evicted before the commit, \Attend asks the MC to redirect and persist them into in-pmem log entries and {\em commits} them \textit{off-chip} upon committing the transaction. \Attend lazily conducts pmem writes only for cache replacements at runtime. On a crash, \Attend saves metadata and data for active transactions into pmem for recovery. Experiments show that, by using CPU cache for both buffering and logging, \Attend yields much higher throughput and 
	incurs significantly fewer pmem writes than state-of-the-art designs.
\end{abstract}

\section{Introduction}\label{sec:intro}
 
A few companies  
 have shipped
byte-addressable 
{\em persistent memory} (pmem) products  that are  
put on the
memory bus for CPU to load and store data~\cite{pmem:SKHYnix,intel:3dx,pmem:Micron,pmem:Dell,pmem:HPE,pmem:Intel,pmem:everspin2}.
In order to popularize the use of pmem,
Intel and other manufacturers
have gradually upgraded architectural facilities.
Intel introduced 
more efficient cache line flush instructions  
(e.g., {\tt clwb})  
to substitute the legacy {\tt clflush}~\cite{intel:tsx,pmem:programming:Scargall2020,arch:primer-persistency:2022}.
Cache line flush enables programmers to flush modified cache lines to
the {\em persistence domain}, 
in which data can be deemed to be persistent upon  
a power outage~\cite{pmem:programming:Scargall2020,pmem:eADR:VLDB-2020,arch:eADR:PLDI-2021,arch:ASAP-persistence:HPCA-2022}.
The concept of persistence domain was initially linked to
the feature of Asynchronous DRAM Refresh (ADR).
ADR keeps DRAM in self-refresh mode and, more important,
places pmem and the write pending queue (WPQ) of memory controller (MC)
in the persistence domain~\cite{txm:Proteus:MICRO17,txm:LAD:MICRO19,arch:Dolos-secure:Micro-2021}, as
it guarantees  
to flush data staying in the WPQ to 
pmem in case of a power outage. Later
Intel extended ADR as {\em eADR} that  
further 
manages to  
flush all cache lines to pmem 
on a crash \cite{eADR:Intel,arch:BBB:HPCA-2021,arch:stealth-persist:HPCA-2021,pmem:eADR:VLDB-2020,arch:eADR:PLDI-2021}.
As a result,
eADR frees programmers from manually flushing cache lines 
to pmem.
Platforms with  
the eADR feature
are 
commercially available today. 
However,  
eADR
factually builds a  
{\em transient persistence domain},
because the eventual persistence of data buffered
in
WPQ entries and CPU cache lines is made by
an uninterruptible power supply flushing all such data to 
pmem.

The advent of pmem has motivated 
programmers to directly operate with persistent data in pmem. 
It is non-trivial to enable 
the {\em atomic durability}
for in-pmem data regarding
unexpected system failures,  
e.g., a  
power outage. 
For example, inserting a key to a sorted array is likely to 
move existing keys that may span a few cache lines.
Programmers  
 make  
such an insertion into a {\em transaction} 
that
shall be atomically modified as a unit. In other words,
the change of cache lines for 
the insertion
must be done in an all-or-nothing  
fashion.
{If a crash occurs,  
after reboot
involved cache lines should either contain all keys including the new 
 one,
or retain only original keys without any movement.}
{Programmers can use the software logging strategy
to back up data for a transaction.} 
{Whereas, 
software logging is ineffectual.}  
{Firstly, it incurs double 
writes, which  
impair both performance and 
lifetime for pmem products 
 \cite{pmem:PCM:Micro-2009,arch:endurance:ASPLOS-2010,arch:PCM-refresh:IEEE-Micro-2011,arch:SD-PCM:ASPLOS-2015,arch:ReRAM-endurance:TACO-2018,pmem:FlatStore:ASPLOS-2020,arch:PCM:HPCA-2022}.
Secondly, 
it executes extra instructions for logging and consumes more
architectural 
resources, such as  
double WPQ entries and 
 pmem spaces 
for data and log copies.
Thirdly,  
software logging must explicitly enforce
the ordering of persisting log copies before
updating data  
through memory fence instructions (e.g.,
{\tt sfence}) to render the backup copy reliable. 
The eADR helps to avoid cache line flushes
but still necessitates the use of memory fences, 
which are costly for achieving the in-pmem atomic durability~\cite{pmem:orderless:ICCD-2014,txm:transaction-cache:DAC17,pmem:fence-less:Micro-2020,pmem:programming:Scargall2020}.
}

{Computer architects have explored how to
enable the atomic durability in various hardware approaches for 
applications to gain the data consistency with
pmem~\cite{txm:Kiln:MICRO13,txm:ThyNVM:Micro-2015,txm:backend:HPCA-2016,txm:Proteus:MICRO17,txm:ATOM:HPCA-2017,txm:transaction-cache:DAC17,txm:PiCL:MICRO18,txm:ReDU:Micro-2018,txm:Steal-but-No-Force:HPCA18,
	txm:LAD:MICRO19,txm:HOOP:ISCA20,txm:MorLog:ISCA-2020}.
They mostly exploit either a redo or undo log copy, or both, 
for data to be atomically modified in a hardware-controlled transaction. 
Some of them considered persistent CPU caches made of non-volatile memory (NVM) technologies to keep redo log copies~\cite{txm:Kiln:MICRO13,txm:transaction-cache:DAC17}.
Some others 
used in-pmem areas for logging 
and added on-chip redo or undo log buffers, or both, within the cache hierarchy~\cite{txm:backend:HPCA-2016,txm:ATOM:HPCA-2017,txm:Steal-but-No-Force:HPCA18,txm:ReDU:Micro-2018,txm:MorLog:ISCA-2020,txm:HOOP:ISCA20}. Recently
researchers explored the transient persistence domain of limited WPQ entries protected by the ADR feature
 to temporarily hold log or data copies~\cite{txm:Proteus:MICRO17,txm:LAD:MICRO19}. 
}

In this paper, we consider leveraging the eADR-supported CPU cache 
to enable the atomic durability for applications. 
The eADR  
guarantees all cache lines to 
be flushed back to pmem
on a power outage, thereby promising substantial space in numerous megabytes
to 
secure crash recoverability for applications.
Modified data staying in a cache line is a natural 
redo log of the in-pmem copy. However, a normal cache replacement
or the eADR on a power failure writes the cache line back
to its home address. 
If the cache line belongs to an uncommitted transaction, the transaction 
cannot be recovered, as 
the overwrite destroys the original copy. 
We hence
develop {\bf \Attend}
to overcome this challenge
with supportive hardware components and comprehensive transactional protocols.
The main points of \Attend  
are summarized as follows.
\begin{itemize}[leftmargin=3mm]\setlength{\itemsep}{-\itemsep}	
	\item \Attend makes CPU cache be both the working memory and main transaction log. It
	enhances a part of CPU cache lines with \textit{transactional} {\em tags} (TransTags) and 
	manages
	an in-pmem log zone holding transaction profiles and log entries
	for spatial extension and emergency use.
	It also customizes the 
	MC between CPU and pmem
	to handle cache lines
	evicted due to  
	cache replacement or 
	eADR on a power-off.
	\item  	\Attend places data that programmers put in a transaction 
	into cache lines with TransTags. 
	On a transaction's commit,
	\Attend \textit{commits on-chip} 
	the transaction's data buffered in CPU cache by modifying TransTags. 
	For  
	cache lines evicted before the commit,
	\Attend  
	makes the MC
	map and persist them to in-pmem log entries. 
	It 
	keeps their mappings for proper reloading until the commit, at which
	it \textit{commits} them {\em off-chip} by changing their states in the MC.
	Then \Attend silently migrates them to their
	home addresses.  
	\item 	A crash initiates the emergency use of in-pmem log zone.
	With eADR,
	\Attend dumps cache lines with TransTags
	and mappings in the MC into a dedicated area of log~zone.  
	To recover, it  
	discards uncommitted transactions and carries 
	on unfinished data write-backs for committed transactions.
\end{itemize}

\Attend exploits 
CPU cache to log and coalesce data updates. 
Only on cache replacement or power outage will \Attend 
passively flush cache lines, which is in contrast to prior  
works that proactively write undo or redo log copies to pmem for backup.
As a result, 
\Attend both achieves high performance and minimizes pmem writes.
We have prototyped  
\Attend within the gem5 simulator~\cite{sim:gem5} and evaluated it thoroughly
with micro- and macro-benchmarks.
Experimental results confirm that
\Attend well supports ordinary workloads of typical applications
and inflicts the least writes to impact the write endurance of NVM.
{For example, running with prevalent workloads, 
\Attend yields about 89.2\%, 29.2\%, and 51.3\% higher throughput on average
than software logging,  
Kiln~\cite{txm:Kiln:MICRO13},
and HOOP~\cite{txm:HOOP:ISCA20}, while the data \Attend writes to pmem is  
29.8\%, 37.4\%, and 1.4\% that of them, respectively.}

\section{Persistence Domain and Atomic Durability}\label{sec:bg}

\subsection{Persistence Domain}

\textbf{Pmem.}\hspace{0.5ex}
Pmem embraces both byte-addressability and persistency. 
Researchers
have considered building pmem with various memory technologies,
such as 
phase-change memory \cite{pmem:PCM:Micro-2009,pmem:PCM:ISCA-2009,pcm:pcm:micro10,arch:PCM-refresh:IEEE-Micro-2011,arch:PCM:HPCA-2011,arch:SD-PCM:ASPLOS-2015,arch:PCM:ISCA-2016},
spin-transfer torque RAM (STT-RAM) \cite{pmem:everspin2,arch:STT-RAM:DAC-2014,txm:checkpoint-SST-RAM:ICCAD-2014,arch:STT-MRAM:ICCAD-2017,arch:STT-MRAM:ICCD-2017,arch:STT-RAM-ECC:TC-2018}, resistive RAM \cite{arch:ReRAM:HPCA-2015,arch:ReRAM:ISCA-2016,arch:ReRAM-endurance:TACO-2018,arch:ReRAM-PIM:ISCA-2019,arch:RRAM-PIM:Micro-2019}, 
3D XPoint~\cite{intel:3dx,pmem:Intel},
and DRAM backed by flash~\cite{pmem:Dell,pmem:HPE,pmem:SMARTM,pmem:Micron,Btree:NV-Tree:FAST-2015}.  
Applications can 
directly load and store data with pmem
\cite{pmem:Mnemosyne:ASPLOS11,NVM:NVM-Duet:ASPLOS-2014,pmem:NV-Heaps:SIGARCH14,intel:3dx,Btree:NV-Tree:FAST-2015,pmem:DudeTM:SIGARCH17,arch:VANS-LENS:Micro-2020,pmem:Timestone:ASPLOS20,pmem:Optane-study:FAST-2020,arch:stealth-persist:HPCA-2021,pmem:programming:Scargall2020,pmem:survey:acmsurv21,txm:ThyNVM:Micro-2015}.

\textbf{Persistence Domain.}\hspace{0.5ex}
Persistence domain
is  
a region of computer system in which data would not be lost but retrievable 
when the system crashes or power failures occur~\cite{pmem:programming:Scargall2020,eADR:Intel}.
It conventionally includes
disk drives at the secondary storage level.
The advent of pmem brings it up to the memory level.

\textbf{ADR.}\hspace{0.5ex}
The ADR   
further extends
the persistence domain to the WPQ of 
MC~\cite{arch:eADR:PLDI-2021,txm:LAD:MICRO19,txm:Proteus:MICRO17,arch:Dolos-secure:Micro-2021}.
ADR guarantees that data received at the WPQ can be flushed to pmem upon a 
power outage.
Though,
the persistence enabled by ADR is transient, as it is the pmem that eventually makes data persistent.
Also,
CPU cache is still volatile  
and cache lines would be lost on a  
crash. 
Thus,
programmers must explicitly flush data staying in cache lines (e.g., \texttt{clwb}) to pmem.
Flushing data from CPU cache to pmem
is not only synchronized and time-consuming, but is also error-prone and hurts programmability~\cite{pmem:programming-is-hard:apsys17,arch:BBB:HPCA-2021}.

\textbf{eADR.}\hspace{0.5ex}	
{Intel extended ADR as eADR which guarantees to 
	flush all cache lines 
	to pmem in case of a power outage by employing extra power supply} 
\cite{eADR:Intel,pmem:programming:Scargall2020}.
{Alshboul et al.~\cite{arch:BBB:HPCA-2021} proposed BBB that
	employs a battery-backed persist buffer alongside each core's L1D cache}
{and achieves an identical effect as eADR with much less cost.} {They help programmers 
	avoid explicit cache line flushes.}
More important,
they
make   
the multi-level CPU cache hierarchy provide
a transient persistence domain in dozens of megabytes
on top of pmem.

\subsection{Atomic Durability}\label{sec:bg-ad}

The atomic durability, or failure-atomic durability, refers to the
crash consistency of modifying in-pmem data in case of a
crash. 
The insertion with an in-pmem sorted array mentioned in Section~\ref{sec:intro}
is a typical {transaction} programmers would 
define with their desired semantics.  
A transaction must be done 
in an atomic (all-or-nothing) fashion.  
Otherwise, a half-done change may 
leave data in ambiguity or uncertainty
after a crash.

\begin{figure}[t]
	\begin{center}
		\includegraphics[width=\columnwidth]{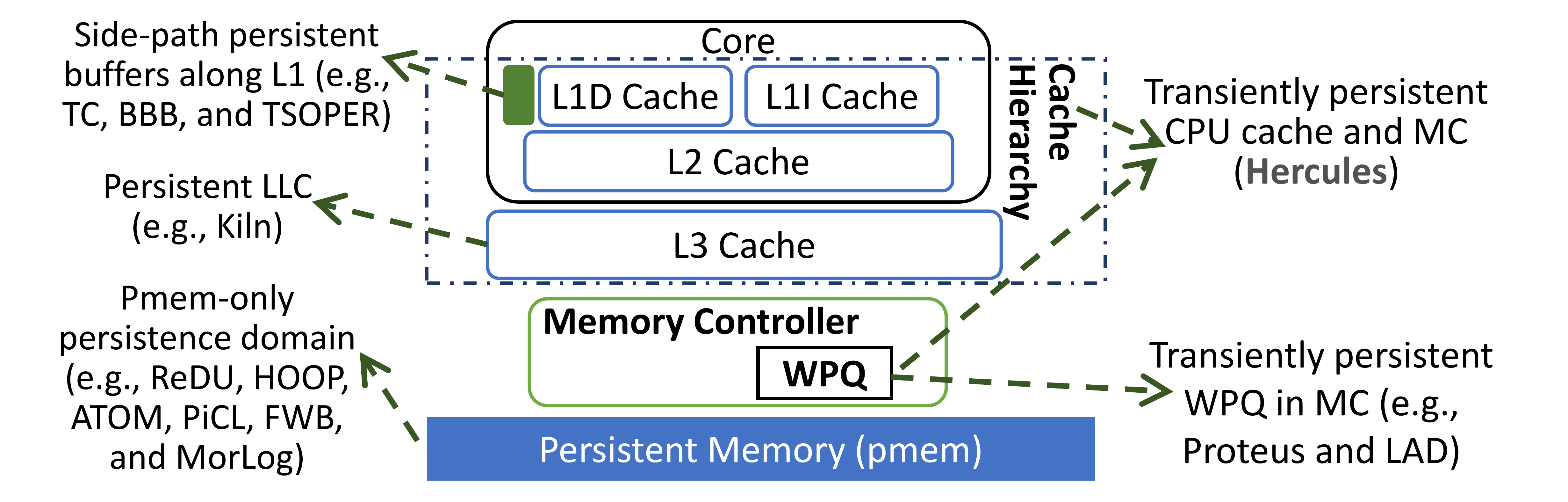}
		\vspace{-1ex}
		\caption{{An Illustration of Persistence Domains Explored in Prior Works and This Paper}} \label{fig:cache-persistency}
	\end{center}
	\vspace{-5ex}
\end{figure}

\subsubsection{Software Solution} 
\ \newline
\indent Modern 64-bit CPUs allow an atomic write of up to  
8 bytes.
Programmers bundle multiple data operations 
for a task  
in one transaction and seek software or hardware solutions.
Software logging is a common technique. Programmers  
explicitly record original (resp. modified) data in an undo (resp. redo) log.
However, 
software logging is not effectual  
with several factors.
Firstly, logging incurs double writes  
due to writing both log and data copies~\cite{txm:Kiln:MICRO13,pmem:UBJ:FAST-2013,txm:ReDU:Micro-2018,arch:PS-ORAM:ISCA-2022}.
Double  
writes 
jeopardize performance and impair lifetime for NVM technologies 
that
have limited write endurance~\cite{pmem:PCM:Micro-2009,arch:endurance:ASPLOS-2010,arch:PCM-refresh:IEEE-Micro-2011,arch:SD-PCM:ASPLOS-2015,arch:ReRAM-endurance:TACO-2018,pmem:FlatStore:ASPLOS-2020,arch:PCM:HPCA-2022}.
Secondly, logging demands extra instruction to be executed.
Log and data copies also consume more
architectural resources.  
For example,
they need double locations in pmem.
If programmers use CPU cache to buffer them, they take double cache lines.
Thirdly, the ordering of writing log copies prior to data must 
be retained by using memory fences,
the cost of which, albeit the presence of eADR, 
is essential and
substantial~\cite{pmem:orderless:ICCD-2014,txm:transaction-cache:DAC17,pmem:fence-less:Micro-2020}.

\subsubsection{Hardware Designs}\label{sec:ad:hardware}
\ \newline
\indent  The essence of gaining atomic durability 
is to make a backup copy before  
modifying data in place.
In order
to back up data,
state-of-the-art hardware designs explore different persistence domains, which categorize
them into three classes.

\textbf{Persistent CPU cache.}\hspace{0.5ex}
{As shown in~\autoref{fig:cache-persistency},
	Kiln works with a persistent  
	last-level cache (LLC)}~\cite{txm:Kiln:MICRO13}.
The persistence domain covers LLC and pmem. 
Kiln manages redo log copies in LLC to back up
in-pmem data.
Later,
Lai et al.~\cite{txm:transaction-cache:DAC17} employed a side-path persistent
transaction cache (TC) along L1 
cache in each CPU core. TC is similar to the 
persistent buffers used in other works~\cite{arch:BBB:HPCA-2021,arch:TSOPER:HPCA-2021}.
Modified cache lines of a transaction
are first-in-first-out (FIFO)
put in the TC  
and serially written 
to pmem 
on committing a transaction.

\textbf{Pmem.} {A few hardware designs were built on
	a pmem-only persistence domain.}
Doshi et al.~\cite{txm:backend:HPCA-2016} proposed to use a victim cache
to hold evicted cache lines that would be 
subsequently written to an in-pmem redo log. 
These cache lines
are eventually 
written to their home addresses
by copying log entries via non-temporal stores.
Similarly, Jeong et al.~\cite{txm:ReDU:Micro-2018} proposed ReDU that utilizes
a DRAM cache to hold evicted cache lines from the LLC.
ReDU directly
writes modified data from DRAM cache
to home addresses, as it
installs a log buffer alongside L1 cache to collect 
and write the in-pmem redo log.

Cai et al.~\cite{txm:HOOP:ISCA20}
designed HOOP  
with a physical-to-physical address indirection layer in the MC, which
helps it
write modified data to a different pmem address for backup
and later  
move data to home addresses.  
Joshi et al.~\cite{txm:ATOM:HPCA-2017} noted that 
the MC loads a cache line for a write request 
and proposed ATOM to 
write the loaded copy to an in-pmem undo log in a parallelized manner.
Nguyen and Wentzlaff~\cite{txm:PiCL:MICRO18} proposed PiCL that also uses the idea of
undo logging with  
an on-chip log buffer. 
PiCL makes a trade-off between performance and durability
by snapshotting and saving data
in an epoch-based periodical checkpointing manner~\cite{txm:checkpoint-SST-RAM:ICCAD-2014}.
Ogleari et al.~\cite{txm:Steal-but-No-Force:HPCA18} and Wei et al.~\cite{txm:MorLog:ISCA-2020}
both chose the undo+redo logging approach.  
Ogleari et al. captured data's 
redo and undo log copies from the in-flight write operation and the write-allocated 
cache line, respectively.
They also used a force write-back (FWB) mechanism to control pmem writes.
Wei et al. studied data encoding with hardware  
logging
so as to only record necessary changes for a transaction's data, thereby reducing pmem writes.
Both designs 
add on-chip undo and redo log buffers.

\begin{figure*}[t]
	\begin{subfigure}{0.193\textwidth}
		\includegraphics[width=\textwidth]{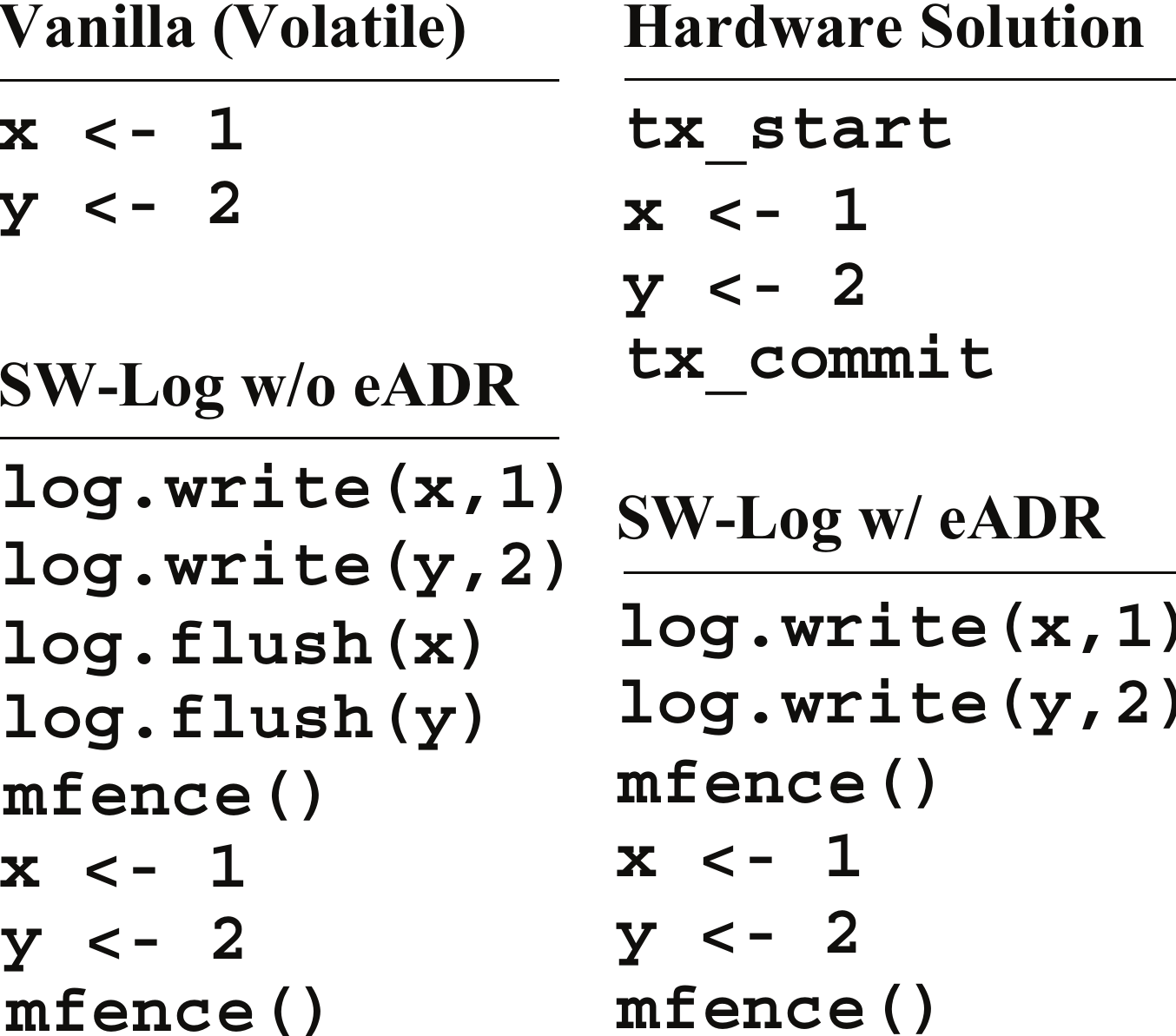}
		\caption{Sample Code on Logging~\cite{arch:blurred-persist:TOS-2016,txm:transaction-cache:DAC17,txm:ATOM:HPCA-2017,txm:Proteus:MICRO17,txm:Steal-but-No-Force:HPCA18,txm:ReDU:Micro-2018}}
		\label{fig:mot:code}
	\end{subfigure}	
	\hfill
	\begin{subfigure}{0.193\textwidth}
		\includegraphics[width=\textwidth,page=1]{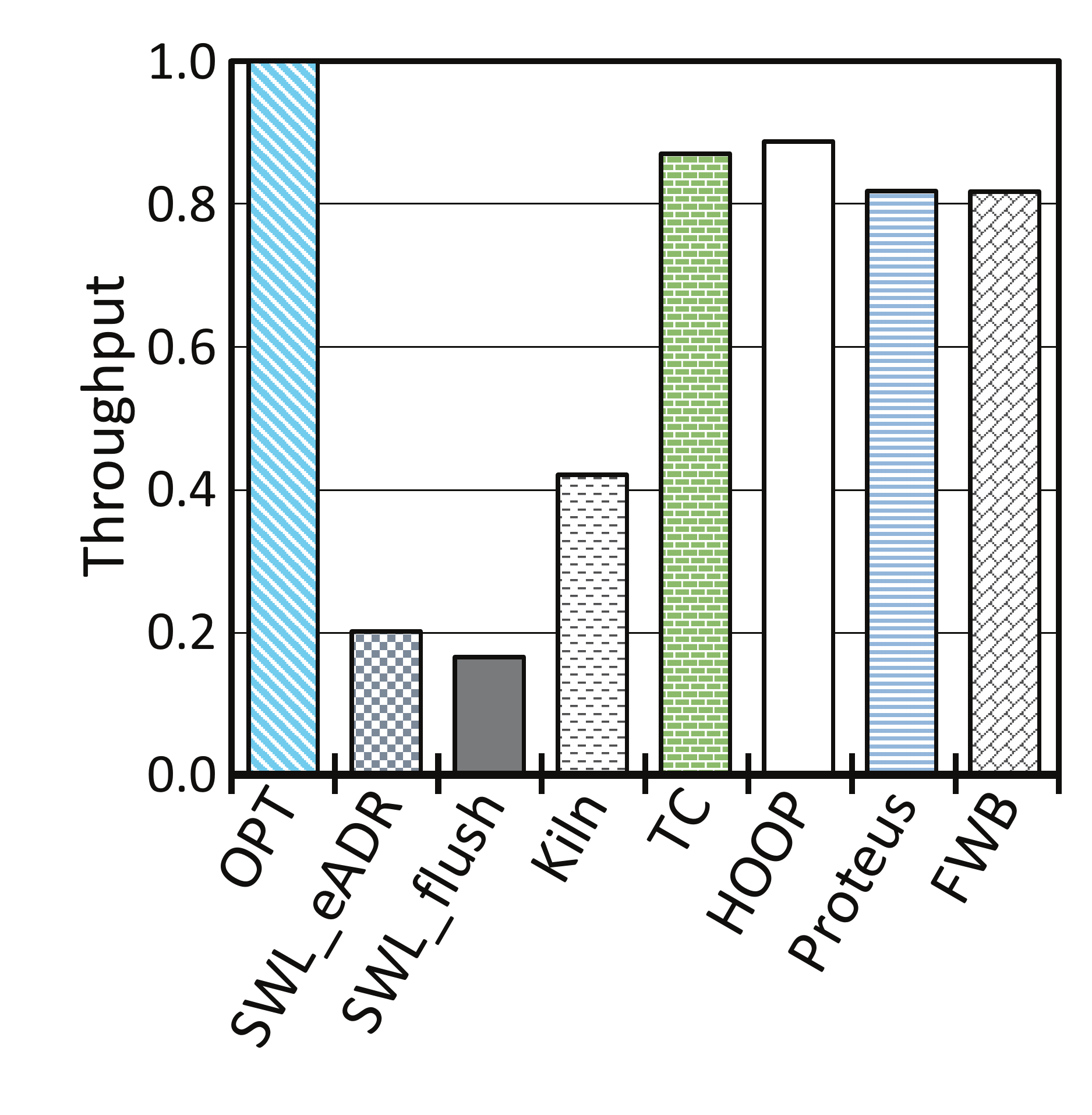}
		\caption{Throughput}
		\label{fig:mot:thruput}
	\end{subfigure}
	\hfill
	\begin{subfigure}{0.193\textwidth}
		\includegraphics[width=\textwidth,page=2]{./mot.pdf}
		\caption{Pmem Writes}
		\label{fig:mot:writes}
	\end{subfigure}
	\hfill
	\begin{subfigure}{0.193\textwidth}
		\includegraphics[width=\textwidth,page=3]{./mot.pdf}
		\caption{Cache Accesses}
		\label{fig:mot:cache}
	\end{subfigure}
	\hfill
	\begin{subfigure}{0.193\textwidth}
		\includegraphics[width=\textwidth,page=1]{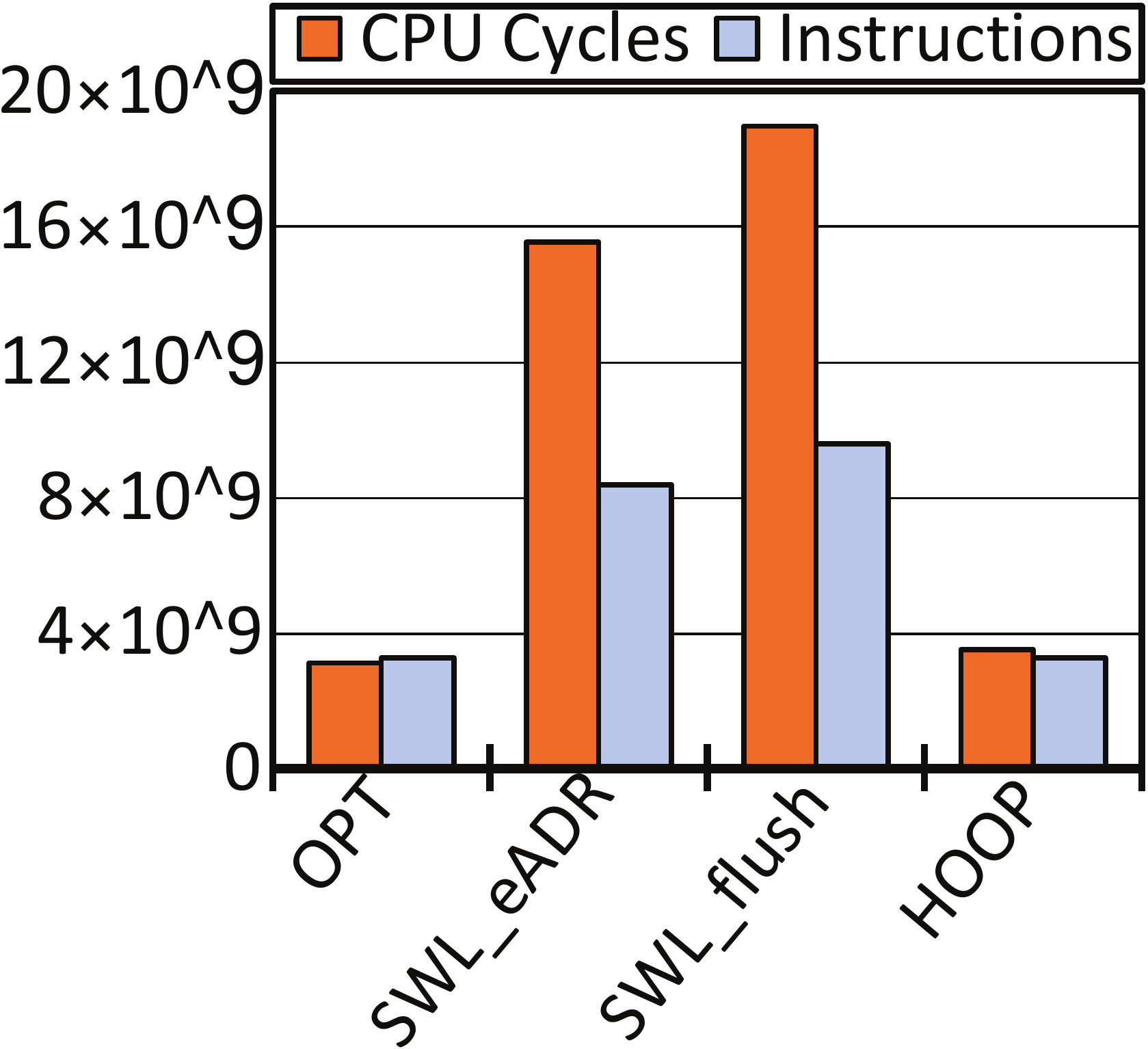}
		\caption{Executed Instructions and CPU Clock Cycles}
		\label{fig:mot:inst}
	\end{subfigure}
	\caption{A Study on Software Logging Logging with/without eADR and State-of-the-art Hardware Designs for Atomic Durability}
	\vspace{-3ex}
\end{figure*}

\textbf{ADR-supported transient persistence domain.}\hspace{0.5ex}
{The ADR places 
	the WPQ of MC in the transient persistence domain}.  
Shin et al.~\cite{txm:Proteus:MICRO17} designed Proteus
that  
considers the  
WPQ to keep log copies.
When a transaction commits, Proteus discards relevant log copies in the WPQ
and hence reduces pmem writes. 
Gupta et al.~\cite{txm:LAD:MICRO19} proposed LAD that also
leverages the ADR-supported 
MC as a staging buffer
to accumulate data updates before committing a transaction.
Whereas,  
for Proteus and LAD,
the limited capability of WPQ  
entails a high likelihood 
of falling back to the use of an in-pmem log.

\section{Motivation}\label{sec:motivation}

{We consider leverage the transient persistence domain in CPU caches} {made by eADR and BBB}
{to enable atomic durability.
We have conducted a motivational study to analyze  
the potential gain introduced by using CPU cache to process transactions.}
\autoref{fig:mot:code} captures four sample code snippets, 
i.e., a vanilla `volatile' program without  
guarantee of atomic durability, 
two  
software redo logging versions {\em without} and {\em with} the eADR 
(both using CPU cache to buffer log copies),
and one version using a typical hardware transactional design.  
Accordingly we have 
tailored B+-Tree
with the volatile version for the optimal performance (OPT),
two versions of software logging (SWL\_eADR and SWL\_flush),
and five prior  
hardware designs (see \autoref{fig:mot:thruput}).
We run them with 
gem5~\cite{sim:gem5} to
insert one million key-value (KV)
pairs (8B/8B  
for K/V).  
We set each insertion as one transaction.
Section~\ref{sec:eval} would detail
our  
evaluation setup and methodology.
\autoref{fig:mot:thruput}, \autoref{fig:mot:writes}, \autoref{fig:mot:cache}, and \autoref{fig:mot:inst} 
show a quantitative comparison on the throughputs normalized against that of OPT, 
the quantity of pmem writes,
and other CPU execution results, respectively.
We can obtain three  
observations from these diagrams.

\textbf{\textfrak{O1}: The eADR  
improves the performance of software logging
and using CPU cache significantly reduces pmem writes}.
Comparing SWL\_eADR to 
SWL\_flush in~\autoref{fig:mot:thruput} tells that
the avoidance of cache line flushes  
makes SWL\_eADR gain
\fpeval{round((0.064537952 - 0.052903335) / 0.052903335  * 100, 1.0)}.0\% 
higher throughput. 
This performance improvement
justifies the usefulness of transient persistence domain for atomic durability
in the software logging approach.
Moreover, as shown in~\autoref{fig:mot:writes}, except Kiln that employs a large persistent LLC for buffering,
the quantity of data written by SWL\_eADR is 
\fpeval{round(1453802048 / 2261137664  * 100, 1)}\%, 
\fpeval{round(1453802048 / 8671028160  * 100, 1)}\%, 
\fpeval{round(1453802048 / 8869065600 * 100, 1)}\%, and
\fpeval{round(1453802048 / 11229635200  * 100, 1)}\%
that of   
TC, 
FWB, Proteus, 
and HOOP, respectively. 
The reason is twofold.
Firstly, the eADR-supported CPU cache in sufficient megabytes
holds both log and data copies over time, so 
SWL\_eADR substantially 
brings down data to be written to pmem.
Secondly, hardware designs mostly need to
write 
backup copies to pmem
for crash recoverability, 
because they  
have been developed without a 
transiently persistent CPU cache hierarchy.
Therefore,
hardware designs generally incur much more pmem writes than SWL\_eADR.

{\bf \textfrak{O2}: 
{Compared to software logging with eADR, 
hardware designs gain higher performance without the
use of eADR, which indicates the potential of a new hardware design
utilizing extensive CPU cache for
atomic durability}}.
{Double writes make a crucial innate
defect for software logging.  
As shown in~\autoref{fig:mot:thruput},
despite no explicit flush of data 
with eADR,
SWL\_eADR is still inferior to hardware designs.} 
{Without loss of generality, we take
HOOP as a representative for illustration.
\autoref{fig:mot:cache} and~\autoref{fig:mot:inst} capture the accesses 
to L1D/L2/L3 caches 
and the number of instructions and clock cycles,
for OPT, SWL\_flush, SWL\_eADR, and HOOP, respectively.} 
{SWL\_eADR underuses the eADR-supported CPU cache,
	incurring}
\fpeval{round((4437079723 - 1241304814) / 1241304814  * 100, 1)}\%, \fpeval{round((40893504 - 21949352) / 21949352  * 100, 1)}\%, and \fpeval{round((37391687 - 17357741)/ 17357741  * 100, 1)}\% 
{more loads and stores  
	to L1D, L2, and L3 caches than HOOP, respectively}.
{HOOP conducts 
address indirection in the MC
for hardware-controlled out-of-place backups. Consequently, it  
performs  
backup operations without using CPU cache and
 achieves comparable cache accesses and instructions 
against OPT}. 
{Due to the	
unawareness of CPU cache used as an ample transient persistence domain, 
 hardware designs like HOOP, Proteus, and FWB
must directly write data into pmem for backup or rely on
 limited WPQ entries.}
{To sum up, SWL\_eADR wastes valuable cache space 
	despite the boost of eADR
	while  
prior hardware designs did not foresee transiently persistent CPU caches.}

{\bf \textfrak{O3}: The eADR promises an ample transient persistence domain and shall be well utilized to achieve the 
				atomic durability, high performance, and minimum pmem writes}.
			As shown in~\autoref{fig:mot:thruput}, 
			an evident gap still exists between OPT and hardware or software designs. 
			The eADR-supported CPU cache is certainly a promising feature with 
			a transiently persistent space in dozens of megabytes. 
			{As justified by our test results,  
			SWL\_eADR does not make the most out of it, while 
			no hardware design has ever exploited it. 
			Kiln, one using  
			STT-RAM as the persistent LLC,
			implicitly manifests
			the potential of eADR-supported CPU cache.}
			The throughput of Kiln is not high, partly because of the
			slower access latency of STT-RAM compared than that of SRAM 
			(see~\autoref{fig:mot:thruput}). Yet due to
			the higher density of STT-RAM,
			Kiln's LLC can absorb more pmem writes (see~\autoref{fig:mot:writes}).
			Additionally, platforms with the eADR feature
			are commercially available today, while
			STT-RAM-based cache is being under development.

			These observations motivate us to consider how to utilize the eADR-supported 
			CPU cache
			when developing a hardware design 
			to efficiently guarantee the atomic durability for applications.
			{A modified cache line and its 
			in-pmem copy naturally form a pair of 
			redo log and  
			backup copies, which 
			implies an opportunity for hardware logging. 
			However, 
			the very nature of transient persistence alludes a challenge. Let us
			assume that we directly use the transiently persistent CPU cache to make a 
			redo log. 
			In case of a 
			cache replacement or power outage, 
			the 
			eADR writes a cache line back to its home address.
			For data belonging to an  
			uncommitted transaction, the write-back  
			destroys the intact backup copy in pmem 
			and renders the transaction unrecoverable.
			As a result,
			to achieve  atomic durability, we need to ensure that cache lines of an uncommitted
			transaction should be written elsewhere on write-backs.   
		Also,
			we shall make the most out of CPU cache   
			to simultaneously
			hold data and log copies for minimizing pmem writes.}
			These
			summarize \Attend' main
			tactics and aims.

\section{The Design of Hercules}\label{sec:design}

\begin{figure*}[htb]
	\centering
	\begin{subfigure}[t]{0.65\columnwidth}
		\centering
		\includegraphics[width=\columnwidth]{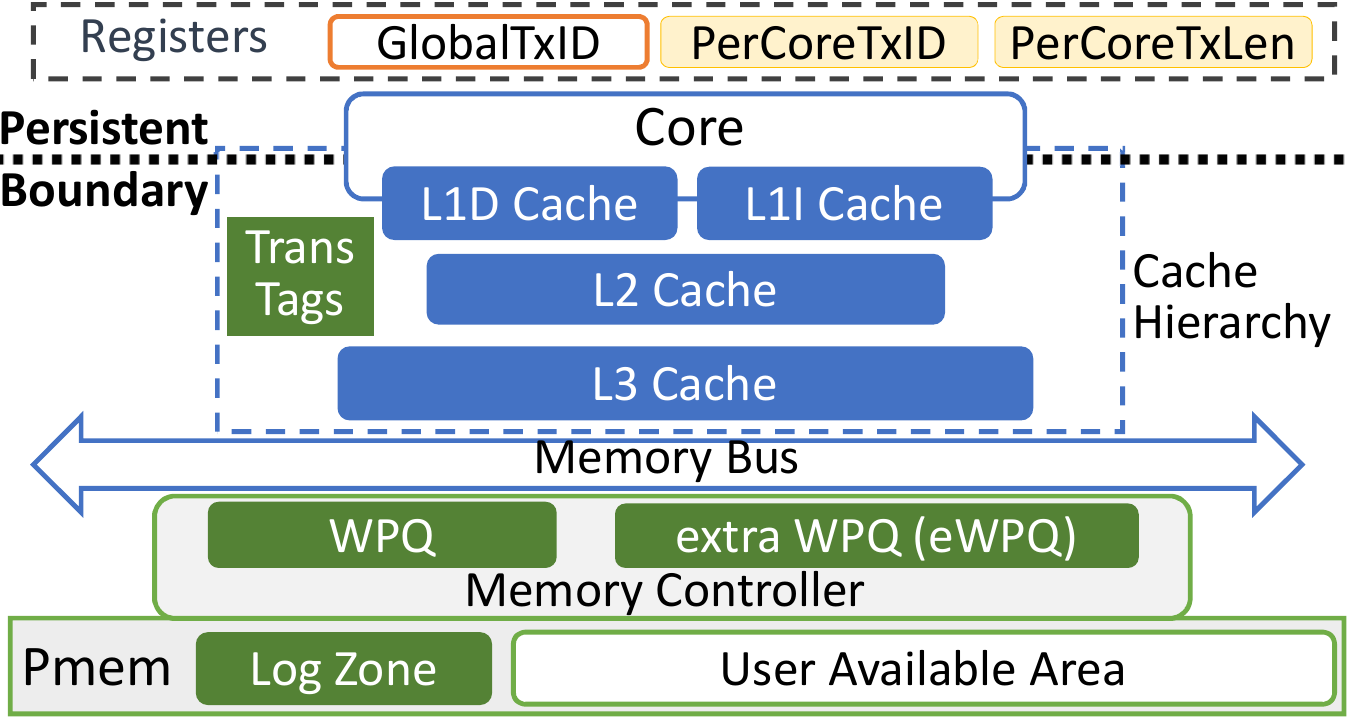}
		\caption{An Overview of \Attend' Components}
		\label{fig:overview}
	\end{subfigure}
	\hfill
	\begin{subfigure}[t]{1.35\columnwidth}
		\centering
		\includegraphics[width=\columnwidth]{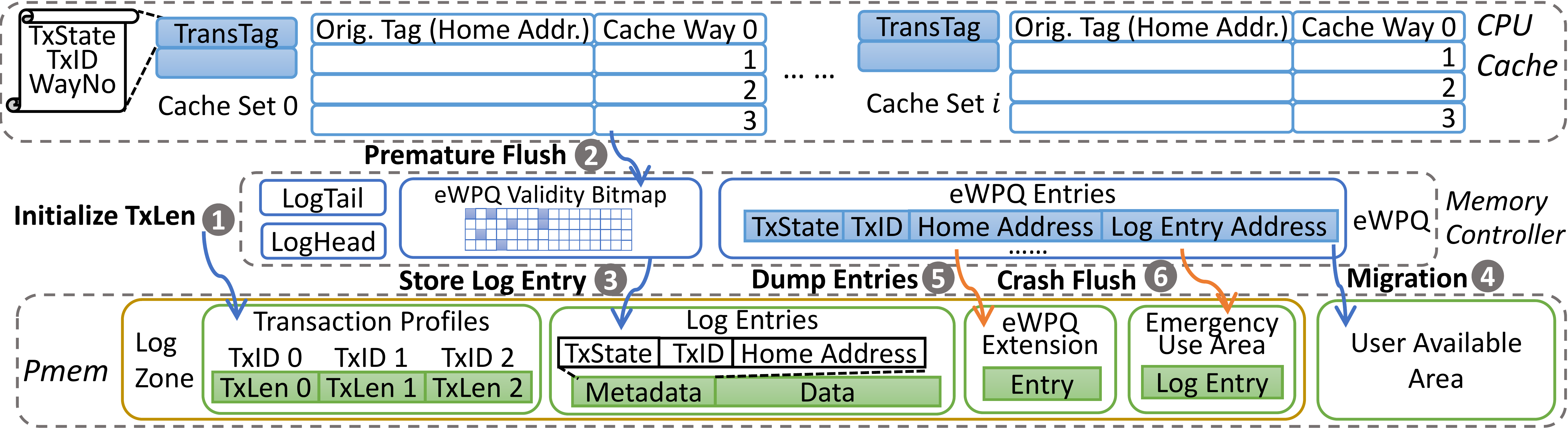}
		\caption{{Detailed Data-paths and Operations of \Attend}}
		\label{fig:detail}
	\end{subfigure}
	\vspace*{-1ex}
	\caption{{An Illustration of \Attend' Components and Operations}}
	\label{fig:Attend}
	\vspace*{-3ex}
\end{figure*}

{\textbf{Overview.}\hspace{0.5ex} 
\Attend   
makes a transiently persistent CPU cache hierarchy
function both as  
working memory and  
hardware-controlled
redo log. 
It installs {\em transactional tags} (TransTags) to a part of cache lines to
hold data for transactions. When a transaction commits,
\Attend \textit{commits} data tracked by TransTags \textit{on-chip}
to reduce pmem writes.  
On evicting cache lines of an uncommitted transaction to pmem,
it places them in an in-pmem log zone rather than their home addresses
to avoid overwriting original data. 
It manages and {\em commits} them {\em off-chip} upon a committing request
through managing an extended WPQ (eWPQ) in the MC.
With a suite of self-contained transactional protocols, 
\Attend efficiently achieves atomic durability with minimized pmem writes
and  
collaboratively works
with other architectural mechanisms.}

\subsection{\Attend' Hardware Components}\label{sec:component}
\Attend uses components distributed in CPU cache, MC, and pmem
to jointly control  
the procedure of transactions  
and manage the versions of data  
for each transaction.
\autoref{fig:overview} captures the main components of \Attend.

\textbf{TransTag.}\hspace{0.5ex} 
A transaction of \Attend is a contiguous series of data operations covering one or multiple cache lines.
In order to manage a transaction's data, 
\Attend adds TransTags  per cache set.
A TransTag has {\tt WayNo}, {\tt TxID}, and {\tt TxState}.
The {\tt WayNo} indicates which actual cache line in the cache set
\Attend is using for a transaction that is 
identified with a unique 21-bit {\tt TxID}.
The reason why 21 bits are used 
is twofold. Firstly,
\Attend demands extra costs such as 
energy, transistors, and wires  
to enable the atomic durability.  
More than 21 bits per {\tt TxID} 
may further increase such costs 
and entail more challenges to achieve systematic efficiency and reliability
(see  
Sections~\ref{sec:energy}). 
Secondly, 21 bits support 
up to $2^{21}$ ($\ge$2 million) 
transactions, which can
satisfy the needs of 
typical applications at runtime.

The {\tt TxState} in one bit shows if the transaction is committed or not.
We call a cache line being occupied by an uncommitted transaction  
{\em transactional cache line},
of which the {\tt TxState} is `1'. Otherwise, {it is non-transactional without a TransTag or
with {\tt TxState} being `0'}. 
In short,
{we configure a logical part of a cache  
set by filling {\tt WayNo}s
for the use of \Attend,
without fixing some cache lines
for 
 transactions}. 
This brings in spatial efficiency and flexibility.
{Given a small amount of transactional data,
	non-transactional data can freely take cache lines
	with {\tt TxState}s unset. Also,
	non-transactional and transactional
data can be placed without swaps}. 
For the illustrative example in \autoref{fig:detail},  
half cache lines of a four-way set ({\tt WayNo} in two bits)
are usable
with two TransTags that cost  
$2\times(21 + 2 + 1) = 48$ bits per set.

{ 
	\Attend  
	manages and uses
TransTags in a similar form   
of the directory used to track 
cache lines	for coherence~\cite{arch:directory:CF-2010,arch:haswell-coherence:ICPP-2015,arch:Multicore-arch,arch:coherence-channel:HPCA-2018,arch:attack-dir:SP-2019}.
It places TransTags alongside L1D and L2 caches in each core and
has a joint bunch of them
for the shared LLC.}
{This organization accelerates filling and changing fields in them.}
{The ratio of TransTag per cache set is defined as the percentage of
TransTags installed over all cache ways of a set.}
As suggested by~\autoref{fig:mot:cache},
we shall consider a higher TransTags ratio for more usable space 
to caches that are closer to CPU,
as they serve more transactional requests. 
The ratio of TransTags  
is thus practically decided by a cache's proximity to CPU.
In practice,
we allot TransTags to all 
cache lines of L1D  
cache.  
The ratio can be half or a quarter for L2 and L3 caches.
We have a 
discussion on this ratio 
in Section~\ref{sec:ratio-eval}.

\textbf{Registers.}\hspace{0.5ex}
\Attend adds a register  
{\tt GlobalTxID} that   
is monotonically increasing, shared by all cores  
to compose the next transaction ID. 
\Attend installs {\tt PerCoreTxID} and {\tt PerCoreTxLen} to each CPU core, holding
the current running transaction's ID and length (number of transactional cache lines), respectively.
These two are critical for the context switch between threads and
stay dormant for threads doing non-transactional operations.
A register  
has 64 bits and 
\Attend 
uses the lower 
bits only 
for extension.
They are kept volatile,
not saved in pmem 
on a crash  
(see Section \ref{sec:recovery}).

\textbf{eWPQ.}\hspace{0.5ex}
Cache lines are evicted over time.
\Attend regularly writes back non-transactional ones  
but  
handles transactional ones specifically in order not to harm in-pmem original copies.
It adds two registers and
an extended WPQ ({\em eWPQ}) along  
WPQ in the MC.
{\tt LogHead} records the address of next available log entry in the in-pmem log zone
while {\tt LogTail} points to the last valid log entry 
(see Section \ref{sec:discussion}). 
On 
evicting a transactional cache line,
the MC allocates a log entry to put data of the cache line
by atomically fetching and increasing 
{\tt LogHead}. 
\Attend 
retains that cache line's
metadata in an eWPQ entry, including {\tt TxID}, {\tt TxState}, 
and and the mapping from home address to 
log entry's address.

\textbf{Log zone.}\hspace{0.5ex}
\autoref{fig:detail} shows  
three areas of in-pmem 
log zone.
The first area (transaction profiles) is  
an  
array of transaction lengths ({\tt TxLen}s)
indexed by {\tt TxID}s. 
\Attend keeps the runtime length of a transaction in
the corresponding thread's context
while it only allows 
two legal values for the in-pmem {\tt TxLen}, i.e., 
an initial zero 
and a non-zero eventual length.  
It uses 
the atomic change of {\tt TxLen} to be non-zero to
mark the commit of a
transaction (see Sections~\ref{sec:transaction} and~\ref{sec:recovery}). 
Using a transaction's 
{\tt TxID} 
to index and find the transaction's
{\tt TxLen} is fast
and brings about lock-free parallel loads/stores for concurrent transactions. 
We set a {\tt TxLen} in 4B,  
so a transaction can cover 
up to $2^{32}$ ($\ge$4 billion) cache lines.

{In the second area, a log entry keeps a cache line evicted from the eWPQ with its data and metadata 
including the home address and {\tt TxState}.} 
{The third area of eWPQ extension receives eWPQ entries evicted at runtime.}
The next area of emergency use is used upon an unexpected crash.  
We show a contiguous log zone in~\autoref{fig:detail}
while it
can be partitioned to 
support concurrent accesses with distributed MCs~\cite{txm:LAD:MICRO19}.

\subsection{\Attend' Transaction}\label{sec:transaction}

\textbf{Primitives.} 
{Like prior works~\cite{txm:Kiln:MICRO13,txm:ATOM:HPCA-2017,txm:Proteus:MICRO17,txm:Steal-but-No-Force:HPCA18,txm:HOOP:ISCA20,txm:transaction-cache:DAC17},
	\Attend has three  
	primitives for programmers to proceed a transaction}, i.e., 
{\tt tx\_start}, {\tt tx\_commit}, and {\tt tx\_abort} to respectively start, commit, and abort a transaction in applications.

Let us first illustrate how \Attend proceeds a transaction  
in an {\em optimistic} 
situation,
i.e., 1) the transaction manages to commit, 2) 
all cache lines of the transaction stay in CPU cache until the commit, i.e., with no cache line %
evicted in  
the entire course of transaction, 
3) \Attend can find a cache line with 
TransTag
available whenever needed,
and 4) no crash   
occurs.

\label{sec:opt}
\textbf{Optimistic procedure.}\hspace{0.5ex}
On a {\tt tx\_start}, 
\Attend atomically fetches a {\tt TxID} from the  
{\tt GlobalTxID} and increments the register by one for subsequent transactions. 
\Attend finds its entry with {\tt TxID}
in the array of transaction 
profiles  
and initialize the {\tt TxLen} 
to be zero (\mininumbercircled{1} in~\autoref{fig:detail}).
Until {\tt tx\_commit} is encountered,
\Attend manages and processes cache lines for data that programmers put 
in the transaction. 
It adopts the write-allocate caching policy.  
Modifying data causes a cache miss
if the data is not in CPU cache and \Attend allocates
a cache line with TransTag before loading the cache line from pmem.
It then fills {\tt TxID}, {\tt WayNo}, and data and
sets the {\tt TxState} as `1'. 
If data already stays in a clean cache line, \Attend 
obtains and configures a TransTag.
{Given a dirty cache line which might be originally non-transactional 
	or belong to a committed transactions,  
\Attend sends that to the next level in the memory hierarchy, e.g., L1D$\rightarrow$L2,
before getting a TransTag, in order not to taint  
 the latest 
  update. 
{When the transactional cache line is evicted to the lower level,
\Attend first sends the older dirty 
non-transactional version to the next lower-level cache or pmem.}
On a crash, 
the older version 
is firstly persisted to the home address.
The upper-level transactional cache line, if committed,
 refills the home address; if not, \Attend writes
it to the area of
emergency use (to be presented).}  
This rules out 
any inconsistency and
uses lower-level caches for  
staging to further
reduce pmem writes.

\Attend follows generic rules in programming transactions. 
It disallows nested or overlapped transactions in one thread, so at most one
transaction is ongoing within a thread. 
{Inspired by Intel TSX~\cite{intel:tsx},
	\Attend provides a configurable option 
to support applications with a fundamental {\em read committed isolation} level \cite{db:mysql:web, db:isolation:SIGMOD95}
to defeat against semantic isolation bugs}.
{Programmers may consider concurrency control mechanisms like locks or semaphores 
between transactions in multi-threading programs.}
More details can be found  
 in Section~\ref{sec:discussion}.

When a thread enters a transaction for the first time, the transaction's  
{\tt TxID} and length are used to fill
 {\tt PerCoreTxID} and {\tt PerCoreTxLen} of the running CPU core, respectively.
{\tt PerCoreTxLen} is incremented by one
every time \Attend 
is going to launch a transactional update
on an uncovered cache line.
If a context switch occurs, the values of  {\tt PerCoreTxID} and {\tt PerCoreTxLen} are 
 saved as
part of the thread's context 
 for an afterward execution.
On committing a transaction,
\Attend atomically sets the in-pmem {\tt TxLen} with 
{\tt PerCoreTxLen} 
and
resets the  {\tt TxState} to be `0' for 
 each transactional cache line
to make it
 visible to other threads.

{The foregoing procedure shows that \Attend
efficiently handles 
a transaction and
\textit{commits on-chip}. 
The eADR 
guarantees committed cache lines would be flushed to
their home addresses in case of a crash.}
Next we present how \Attend 
handles 
conditions not covered in the optimistic circumstance.

\textbf{Premature flush.}\hspace{0.5ex}
The eADR enables \Attend to write back data on 
cache replacements rather than explicit cache line flushes, 
so
updates to a cache line are 
coalesced
and pmem bandwidths are saved.
\Attend handles the write-back of a non-transactional 
cache line with the MC's WPQ in the ordinary way.
For an evicted cache line recorded in a TransTag,  
if {\tt TxState} is `0', i.e., being
non-transactional, 
the MC writes back the cache line to the home address.
If {\tt TxState} is `1', 
\Attend initiates a {\em premature flush} with the MC's eWPQ (\mininumbercircled{2} in 
\autoref{fig:detail}).
As shown in~\autoref{fig:detail}, the 
eWPQ is made of 
eWPQ entries and {an eWPQ validity bitmap
to track the validity status of each eWPQ entry}.
The MC finds a free eWPQ entry for the evicted transactional cache line
and allocates a log entry by atomically fetching and 
increasing the {\tt LogHead}.
The MC 
copies {\tt TxID}, {\tt TxState}, home address, and log entry's address to
the eWPQ entry and asynchronously writes back the transactional cache line in the  
log entry (\mininumbercircled{3} in \autoref{fig:detail}).

\Attend employs the eWPQ 
both for logging  
uncommitted data  and loading
proper
data. When a thread  
resumes execution, 
it may use a cache line that has been prematurely 
flushed. 
The cache line may be from any transaction that is already committed
or  this resumed  
transaction.
\Attend  
references
cache lines  
regarding 
their home addresses.
On a load request, the MC 
checks if the target address 
matches any eWPQ entry 
and
simultaneously tests the {\tt TxState}.
{Given a match and  
`1' {\tt TxState}, if the {\tt TxID} is compared to be the same as 
ongoing {\tt TxID}, \Attend gets a potential hit.}
A match with `0' {\tt TxState} is also likely  
a hit,
because 
the cache line had been evicted  
before the relevant transaction
committed, but not migrated to the home address yet.  
No match results in a miss.

Regardless of a hit or miss, once receiving a request,
the MC starts loading the cache line from the home address.
A hit at the eWPQ 
fetches the corresponding log entry and halts the load
from home address.
{MC checks the full address stored in the log entry 
and forwards it to the CPU cache in case of a true match.}
When CPU cache receives the log entry, MC nullifies the matching eWPQ entry.
A miss  
continues the load of cache line from home address.
In addition,  
a forbidden access from an ongoing transaction
may happen to an eWPQ entry with mismatched {\tt TxID} 
and `1' {\tt TxState}.
\Attend 
aborts that transaction with an exception (see Section~\ref{sec:discussion}
for isolation).

The other reason for employing the eWPQ is to
commit and migrate cache lines that have been prematurely flushed.
A transaction may commit without reusing them.
\Attend exploits the eWPQ to deal with them.
There are two ways to deal with such cache lines.
One is to load them into CPU cache for committing on-chip
and store them
to home addresses by cache replacements.
The other one is to  
{\em commit} them {\em off-chip} by resetting the {\tt TxState}s 
in corresponding eWPQ
entries and
migrating them from log entries to
home addresses via non-temporal stores.
We choose the 
second 
way to reduce cache pollution. 
For efficiency, we 
periodically scan eWPQ entries for 
data migrations.  
The period is configurable,  
set to be every three million instructions in our tests.
A completion of migrating a log entry
clears the validity bit for the eWPQ entry 
(\mininumbercircled{4} in \autoref{fig:detail}). 
A load request that happens before the reset of validity bit
still 
fetches data from 
the log entry.

Previous works have justified  
the efficacy and %
efficiency of using a part of address for cache
management~\cite{arch:SHiP:Micro-2011,arch:KPC:ASPLOS-2017,arch:NVM:density:ISCA-2018,arch:cache-ML:HPCA-2021}.
We  
accordingly
devise a compact eWPQ entry that holds one bit of {\tt TxState} and
three fields of 63 bits evenly partitioned for
{\tt TxID}, 
home address, and  
log entry's address. 
{Our evaluation shows that transactions of typical applications
	are empirically small and generally take few to dozens of cache lines (see
	Section~\ref{sec:ratio-eval}), so
	a home address and a {\tt TxID} in overall 42 bits 
	are sufficient for indexing.
If duplicate matches occur, an eWPQ entry leads to a log entry
	that holds the full home address to rule out ambiguity.}
We manage the eWPQ like a fully associative cache and
set a default size of 4KB for 512 entries, which are ample to
serve ordinary workloads found in
typical applications (see Section~\ref{sec:ratio-eval}). 
We believe a larger eWPQ is practically viable~\cite{txm:LAD:MICRO19,arch:Dolos-secure:Micro-2021}.
{Yet we take into account the very low likelihood of a full eWPQ, wherein
	\Attend evicts  
	the least-recently-used (LRU) 
	entries to an in-pmem eWPQ extension area that is ten times larger than the eWPQ  
	(\mininumbercircled{5} in \autoref{fig:detail}).
	If a request misses in the eWPQ,
	\Attend checks the eWPQ extension area with a target home address
	to properly fetch the corresponding log entry.}

\textbf{Flush on a power-off.}\hspace{0.5ex} When a power-off occurs, 
\Attend flushes WPQ, {\tt LogHead}, {\tt LogTail},
eWPQ, and then
all cache lines 
to pmem. 
\Attend writes non-transactional cache lines to 
home addresses. 
In case of a crash,
it dumps transactional ones to the area of
emergency use 
in the log zone with home addresses and TransTags (\mininumbercircled{6} in \autoref{fig:detail}).
These metadata and data are useful for  
recovery (see Section~\ref{sec:recovery}).

\textbf{Cache replacement.}\hspace{0.5ex} 
{Transactional and non-transactional cache lines  
flexibly	share 
a cache set.}
\Attend   
considers an effective algorithm \cite{arch:SHiP:Micro-2011,arch:KPC:ASPLOS-2017,arch:NVM:density:ISCA-2018,arch:cache-ML:HPCA-2021,arch:replacement:HPCA-2022}
to select a victim for eviction, but 
entitles a higher priority to transactional ones for staying in cache. 
A request to 
place a non-transactional cache line only
replaces a 
non-transactional one in the set.   
To place a transactional one, \Attend  
tries to find 
a free TransTag 
and may replace a non-transactional victim. 
If all TransTags are being 
occupied,
\Attend evicts a transactional cache line.

{\bf Transaction abortion.}\hspace{0.5ex}
A transaction may abort due to various events like  
exception, fault, 
or running out of memory.  
On an abortion,
data recorded in the TransTags and eWPQ entries  
with {\tt TxID}s matched and {\tt TxState}s being `1's are invalidated and discarded,
incurring no harm to original  
data.

\subsection{The Crash Recoverability of \Attend}\label{sec:recovery}

\Attend puts a specific flag in the log zone
to mark a normal shutdown or not.
The flag is not set if any transactional cache line 
is saved to the area of emergency use on power-off.
Regarding an unset flag, 
\Attend recovers at the transaction level  in order
to support applications recovering with semantics. 
As modifying {\tt TxState}s  
of multiple cache lines  
cannot be atomic,  
\Attend atomically
sets the {\tt TxLen} to commit a transaction. 
In recovery,
it fetches eWPQ, {\tt LogHead}, 
and {\tt LogTail} into the MC and scans   
transaction profiles.

\Attend discards transactions with zero {\tt TxLen}s.
As to a committed one,
\Attend scans the area of
emergency use 
to find out
cache lines with {\tt TxID}s matched and {\tt TxState}s being `1's.
\Attend moves them to their home addresses.
In addition, some cache lines of the transaction might have been 
prematurely flushed before the commit, being tracked by the eWPQ,
but not migrated yet prior to the power-off.
That explains why \Attend has saved the entire eWPQ.
{If an entry with a matching {\tt TxID}
is  
valid in the eWPQ validity bitmap},
\Attend migrates  
the mapped log entry 
and then clears the corresponding validity bit.
{\tt LogTail} may be 
 moved after 
  the migration. 
Once moving all such cache lines is completed,  
\Attend resets the transaction's
{\tt TxLen} to be zero. This atomic write rules out  
ambiguity if a 
crash takes place in an ongoing 
recovery.
After resetting {\tt GlobalTxID} and clearing the eWPQ,
\Attend is ready to recommence new transactions.

\subsection{Discussion}\label{sec:discussion}

{\bf Granularity.}\hspace{0.5ex} 
A cache line is the unit transferred between CPU cache and memory, so
\Attend chooses it as 
the unit for transactional operations.
Using programmer-defined variables is more fine-grained but
must incur higher cost and complexity.

\textbf{Exclusion of transactional cache lines.}\hspace{0.5ex} 
\Attend has no restriction of inclusion or exclusion
on non-transactional cache lines.
It enforces an  
{\em exclusive} multi-level cache hierarchy
to transactional cache lines, which means 
 an evicted transactional cache line,
once 
reloaded into the higher-level cache,  
will be removed from the lower-level cache.
The reason of doing so  
is  
twofold.
{Firstly, an exclusive housing of transactional cache lines substantially
saves TransTags and cache space.
Secondly, an exclusion management helps to reduce microarchitectural 
 actions.
Given a transactional cache line 
loaded to a higher level, e.g., L2$\rightarrow$L1D,
for read  
purpose,
an inclusive cache hierarchy has it at both levels holding the latest version.
 When the transaction commits,
 \Attend needs to
reset {\tt TxState}s twice for the same cache line across levels.
Given a transactional  
cache line loaded to a higher level for updating, 
cache coherence protocols like MESI or MESIF help an inclusive cache hierarchy
to invalidate the older version at L2 upon a modification,
but that costs a  microarchitecture-level coherence state transition ($M\Rightarrow I$).}
\Attend' exclusion  
circumvents these inessential state resets or transitions.

\textbf{Isolation.}\hspace{0.5ex}
{Not all pmem systems supporting transactions provide
	thread-atomicity (isolation)~\cite{pmem:survey:ACM-survey-2021}.}{Like using prior 
 hardware designs~\cite{txm:Kiln:MICRO13,txm:backend:HPCA-2016,txm:ATOM:HPCA-2017,txm:HOOP:ISCA20,txm:Proteus:MICRO17}, 
with \Attend
programmers are responsible for the isolations between threads via
concurrency control
methods (e.g., locks or semaphores).}
{We consider that 
 an error-prone application may misbehave on isolations and provide
 a configurable option for \Attend. In brief,
following Intel TSX~\cite{intel:tsx},  
\Attend 
aborts a transaction when 
threads contend to modify transactional data.}
Also, uncommitted data is 
 invisible to other threads.
If one thread 
does a read-only operation on 
data updated in an ongoing
transaction, \Attend 
loads the data's original copy with customized non-temporal hint and data-path~\cite{arch:rowhammer-nt:SP-2020,arch:ntload-pmem,arch:non-temporal:POPL-2022}
either 
from  
lower-level caches or pmem, depending on where the data's original 
copy is (see Section~\ref{sec:opt}).
{Thus, \Attend enables an optional support of the
read committed isolation~\cite{db:mysql:web, db:isolation:SIGMOD95} while
software
concurrency control can 
promote higher isolation levels}.

{\textbf{Coherence.}  
	\Attend collaboratively works with cache coherence protocols  
	like MESI or MOESI.
	{TransTags and the directory for cache coherence~\cite{arch:directory:CF-2010,arch:haswell-coherence:ICPP-2015,arch:Multicore-arch,arch:attack-dir:SP-2019} 
		share similarities in use 
		and we can integrate them to jointly track cache lines.}
\Attend 
does not affect the sharing of
non-transactional cache lines.  
As to transactional ones, 
the application's concurrency control and architectural
{\tt TxID}s in TransTags prevent other cores from modifying  
or fetching uncommitted versions of them.
The  
exclusion and isolation of transactional cache lines
also avoid broadcasting invalidation messages to all cores upon committing a transaction.
In all,
\Attend inflicts no harm but reduces state transitions
to maintain cache coherence.}

{\bf State reset.}  \hspace{0.5ex} 
{Because cache lines of a transaction are likely to 
	be scattered in a multi-level 
	cache/memory hierarchy,
	all prior designs
	 commit them with concrete efforts for atomic durability \cite{txm:Kiln:MICRO13,txm:transaction-cache:DAC17,txm:HOOP:ISCA20,txm:Steal-but-No-Force:HPCA18,txm:ATOM:HPCA-2017,txm:ReDU:Micro-2018}.
 For example, 
 TC forcefully persists data in the side-path cache to pmem~\cite{txm:transaction-cache:DAC17} while
 Kiln flushes down cache lines from upper-level volatile caches~\cite{txm:Kiln:MICRO13}.
 HOOP migrates all data staying in its out-of-place (OOP) buffer installed in the MC 
to its in-pmem OOP region~\cite{txm:HOOP:ISCA20}. FWB has to wait for the drain of current log updates~\cite{txm:Steal-but-No-Force:HPCA18}.}
{Comparatively, \Attend' commit is much more efficient and lightweight, as it just
	sets {\tt TxLen} and
resets {\tt TxState}s for cache lines that a transaction covers.}
{Since TransTags form a structure similar to the directory
	for cache coherence}{used  
		to track and transit states for cache lines,
\Attend employs an auxiliary circuit 
to select ones with a {\tt TxID} 
and clear their {\tt TxState}s.}
{For
data that might be prematurely flushed before the commit,
\Attend uses the auxiliary circuit to 
notify the integrated 
MC and wait for the completion of
 resetting {\tt TxState}s in relevant eWPQ entries.}{Generally 
 these resets can be swiftly done like state transitions
 for cache coherence.}
{We preset a uniform state reset latency  
in which \Attend is supposed to finish, 
with
a 
  discussion  
presented  in Section~\ref{sec:scan-eval}}.

\begin{figure}[t]
\begin{center}
	\scalebox{0.82}{\includegraphics[width=\columnwidth]{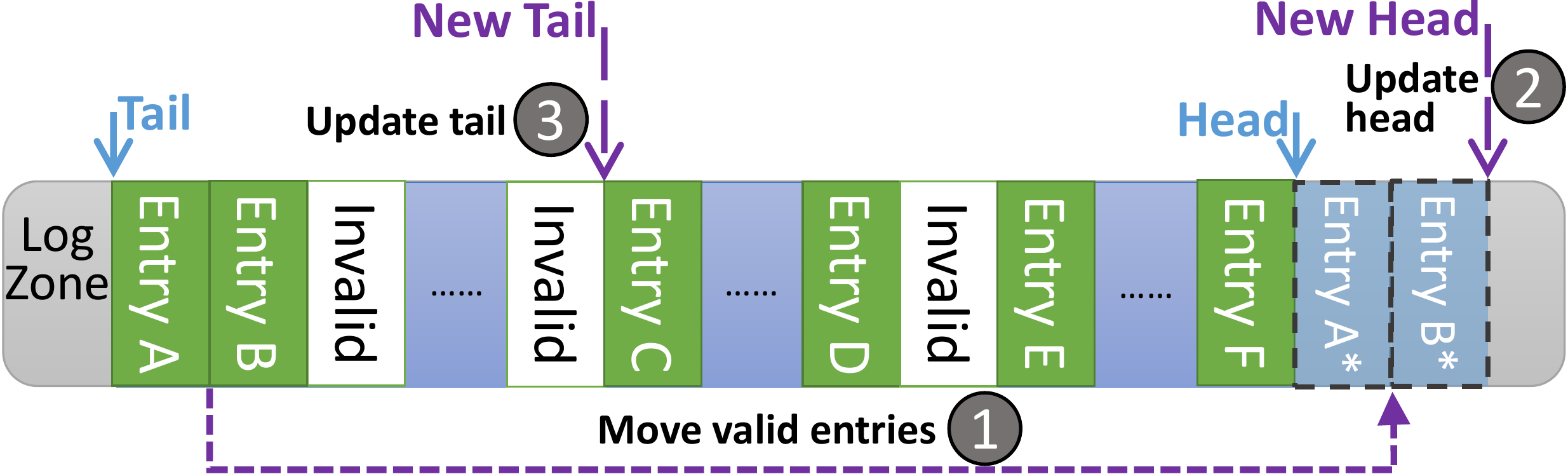}}
	\caption{Garbage Collection on Log Entries.}
	\label{fig:gc_entries}
\end{center}
\vspace{-6ex}
\end{figure}

{\bf Garbage collection (GC) on log entries.}\hspace{0.5ex}
Concurrent transactions 
commit at different time 
 and take up discontinuous  
 log entries.
Committed log entries become  
invalid, scattered across the log zone. 
\autoref{fig:gc_entries} shows  
how \Attend cleans them up.
{\tt LogHead} and {\tt LogTail} frame a window of log entries \Attend is using
while the eWPQ tracks all valid ones.
When {\tt LogTail} has not moved for a while,
numerous invalid entries might accumulate in the window.
If \Attend monitors that the distance between  {\tt LogTail} and {\tt LogHead} is greater than a threshold, e.g., $2^{20}$, it
will initiate a 
GC (\mininumbercircled{1}\mininumbercircled{2}\mininumbercircled{3} in~\autoref{fig:gc_entries}).
\Attend fetches a chunk (e.g., 32)
of successive log  
entries starting at {\tt LogTail} (\mininumbercircled{1}).
{It appends valid uncommitted ones to the locations pointed by {\tt LogHead}} and updates
corresponding eWPQ entries (\mininumbercircled{2}).
Only after updating each eWPQ entry will \Attend move {\tt LogHead} by one.
Then
it slides 
{\tt LogTail} over the chunk
to the next valid 
 entry (\mininumbercircled{3}).
\Attend orderly performs these steps
to preclude any crash inconsistency.
{Note that the system may crash
 in a GC, particularly when a power outage has happened after a movement,
 which results in a moved log entry 
   existing both at {\tt LogHead} and {\tt LogTail}. 
 Such a log entry can be ignored as it belongs to
 an uncommitted transaction with {\tt TxLen} being %
  zero, without any loss of \Attend' crash recoverability.
}

\textbf{Log extension.}\hspace{0.5ex}  
In a very low likelihood,
excessive transactional data might occasionally overfill 
the entire CPU cache in dozens of megabytes 
or even flood the log zone.  
\Attend continues cache placements and replacements 
to swap in and out data, respectively, to proceed transactions. 
We can configure a log zone in gigabytes or even larger. In case that such a large space
is still to be used up, we extend the log zone by using 
the end part of default
log zone to store indirect indexes to the space in a new  
log zone allocated on-demand elsewhere.  
This 
is like the strategy of indirect blocks used by file systems   
to manage big files~\cite{pmem:SCMFS:SC-2011}.  
We also enhance the eWPQ 
with more entries and  
extend 
an eWPQ entry 
with one more bit  
to tell the MC 
if it needs to do indirect references or not to find actual data for
a prematurely
flushed cache line.

\section{Evaluation}\label{sec:eval}

We implement  
\Attend
with gem5  
in the syscall emulation (SE) mode and `classic caches' model. 
\autoref{tab:config} captures the settings of CPU with three-level caches
that align 
with
prior works~\cite{txm:Kiln:MICRO13,txm:transaction-cache:DAC17,txm:Proteus:MICRO17,txm:HOOP:ISCA20,txm:PiCL:MICRO18}. 
We configure an in-pmem log zone in 256MB.
We set default TransTag
 ratios to be 100\%/50\%/25\% for L1D/L2/L3 caches.  
{As L1D cache impacts the most on performance in CPU cache hierarchy,
in order to make a fair and strict evaluation on \Attend,
we  
reduce L1D size   
from 32KB to 30KB for \Attend
by evenly removing some ways in cache sets within gem5 to  
counterbalance the spatial cost of it.}
We further estimate the spatial and energy costs for \Attend in Section~\ref{sec:energy}.
{For a cache line access involving a TransTag,
we increase the tag latency by 30\% as extra time cost.}
{We set the state reset latency  
in ten clock cycles by default}{with a discussion in Section~\ref{sec:scan-eval}}. 
We also
set ten clock cycles for searching the eWPQ  
to check if a cache line has been prematurely flushed.  

\begin{table}[t]
	\centering
	\caption{{System Configuration}}\label{tab:config}
	\vspace{-1ex}	
	\resizebox{\linewidth}{!}{
		\begin{tabular}{|l|l|l|}
			\hline
			Component  & Setting  & Remarks \\ \hline\hline
			\multirow{2}{*}{Processor} & 3GHz, out-of-order, 8 cores, {ROB size=192~\cite{arch:haswell-coherence:ICPP-2015,arch:SIPT:HPCA-2018,arch:reliable:HPCA-2022},} & \multirow{8}{*}{Generic} \\ 
			& {issue/write-back/commit width=8, {MESI protocol} } &\\\cline{1-2}
			L1I Cache & 32KB, 4-way, 2-cycle latency &\\ \cline{1-2}
			L1D Cache & 32KB, 4-way, 2-cycle latency&\\ \cline{1-2}
			L2 Cache & 256KB, 8-way, 8-cycle latency&\\ \cline{1-2} 
			LLC & 16MB, 16-way, 30-cycle latency&\\  \cline{1-2}
			\multirow{2}{*}{Pmem} & Read/write latency=150ns/100ns,  capacity=512GB, & \\ & single channel, read/write buffer size=64&\\  \hline
			Smaller L1D  & 30KB, 4-way, 2/2 cycles read/write latencies&  \Attend \\  \hline			
			Side-path TC&4KB, FIFO, 40/50 cycles read/write latencies&  TC~\cite{txm:transaction-cache:DAC17} \\ \hline
			STT-RAM LLC & 64MB, 16-way, 40/50 cycles read/write latencies& Kiln~\cite{txm:Kiln:MICRO13} \\ \hline
		\end{tabular}
	}
	\vspace{-4ex}
\end{table}

\begin{table}[b]
	\vspace{-3ex}	
	\centering
	\caption{Benchmarks Used in Evaluation}\label{tab:benchmark}
		\vspace{-1ex}
	\resizebox{\linewidth}{!}{
		\begin{tabular}{|l|l|l|}
			\hline
			Category& Benchmark& Remarks \\ \hline\hline
			& Array Swap & Swaps two elements in an array  \\\cline{2-3}
			& Binary Heap & Inserts/deletes entries in a binary heap  \\\cline{2-3}
		Micro-	& B+-Tree & Inserts/deletes KV pairs in a B+-Tree \\\cline{2-3}
		benchmarks	& Hash Table & Inserts/deletes KV pairs in a hash table \\\cline{2-3}
			& Linked List & Inserts/deletes entries in a linked list \\\cline{2-3}
			& RB-Tree & Inserts/deletes KV pairs in a RB-Tree \\\cline{2-3}
			& SDG & Inserts/deletes edges in a scalable large graph \\\hline
			Macro-& TPC-C & OLTP workload (New-order transactions)  \\\cline{2-3}
			benchmarks& YCSB & 80\%/20\% of write/read \\\hline			
		\end{tabular}
	}
\end{table}

\autoref{tab:benchmark} lists 
micro- and macro-benchmarks we use.
{We consider ones that have been widely used in prior works~\cite{txm:Kiln:MICRO13,txm:HOOP:ISCA20,txm:ATOM:HPCA-2017,txm:Steal-but-No-Force:HPCA18,txm:ReDU:Micro-2018,arch:Dolos-secure:Micro-2021}
	and our evaluation methodology strictly follows them.}
We run one million transactions 
  with each benchmark.
 
\begin{figure*}[t]
	\begin{minipage}[b]{1.51\columnwidth}
		\includegraphics[width=\textwidth,page=1]{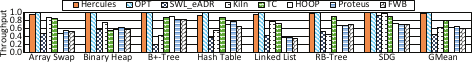} 
	\caption{A Comparison on Throughput of Micro-benchmarks (Normalized against OPT's)}
	\label{fig:micro-txThroughput}
	\end{minipage}
	\hspace{0.01\textwidth}
	\begin{minipage}[b]{0.5 \columnwidth}
		\includegraphics[width=\textwidth,page=1]{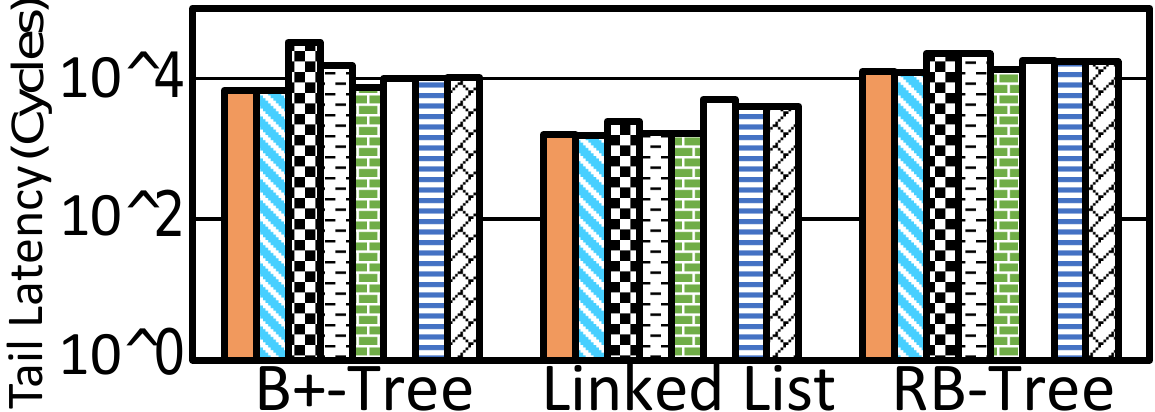}  
		\caption{{99P Tail Latency}} 
		\label{fig:99p}  
	\end{minipage}
\vspace*{-5ex}
\end{figure*}

\begin{figure*}[t]
	\begin{minipage}[b]{1.51\columnwidth}
		\includegraphics[width= \textwidth,page=1]{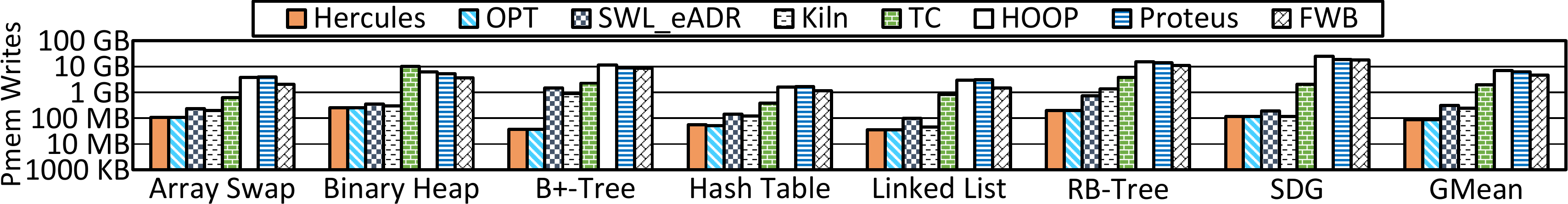} 
	\caption{A Comparison on Pmem Writes of Micro-benchmarks {(Y Axis in Logarithmic Scale)}}
		\label{fig:micro-txwrites}
	\end{minipage}
	\hspace{0.01\textwidth}
	\begin{minipage}[b]{0.5\columnwidth}
		\includegraphics[width=\textwidth,page=1]{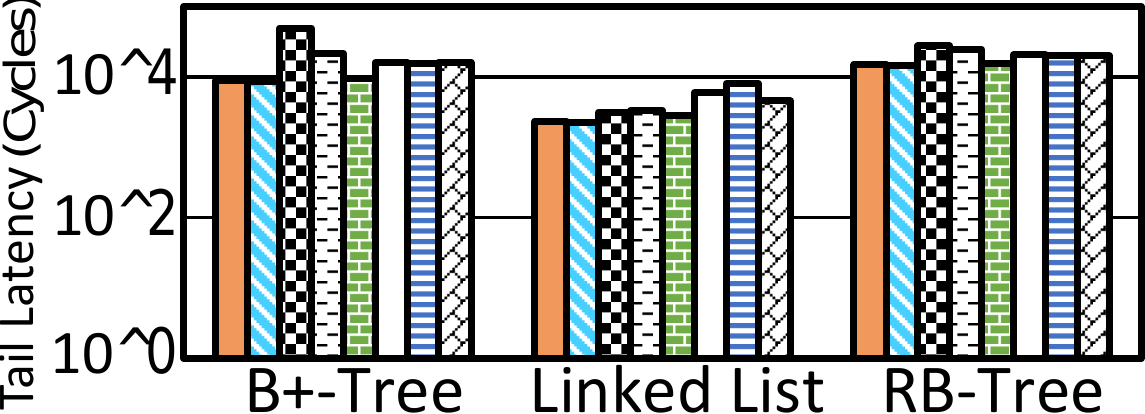}  
		\caption{{99.9P Tail Latency}}
		\label{fig:999p}  
	\end{minipage}
\vspace*{-6ex}
\end{figure*}

We compare \Attend to state-of-the-art hardware designs,
including Kiln, TC, HOOP, Proteus, FWB, and PiCL.
They 
represent different
approaches to guarantee  
atomic durability 
 for in-pmem data 
(see Section~\ref{sec:bg-ad}).
As PiCL is  
with periodical checkpointing,
we discuss it separately
at the end of Section~\ref{sec:micro-eval}.
In~\autoref{tab:config}, 
the STT-RAM LLC and side-path cache are for Kiln and TC, respectively,
both configured in line with
STT-RAM's characteristics 
\cite{txm:Kiln:MICRO13,arch:STT-RAM:ISCA-2011}.
Software
logging with eADR (SWL\_eADR)
is also compared  
 while 
software logging with cache line flushes  
is omitted for brevity.

\subsection{Micro-benchmark}\label{sec:micro-eval}

{\bf Throughput.}  
We normalize the throughputs (txn/$\mu s$, transactions per microsecond)
of all designs against that of OPT. 
As shown in~\autoref{fig:micro-txThroughput}, we
include
the geometric mean of throughputs over benchmarks 
for a high-level  
overview~\cite{txm:Kiln:MICRO13,txm:transaction-cache:DAC17,txm:Proteus:MICRO17,txm:HOOP:ISCA20,txm:PiCL:MICRO18}. 
\Attend achieves comparable performance to OPT.
It significantly  
outperforms prior works
with on average  
89.2\%, 29.2\%, 15.2\%, 51.3\%, 48.0\%, and 57.0\%
higher throughput 
than SWL\_eADR, Kiln, TC, HOOP, Proteus, and FWB, respectively.
\Attend leverages  
CPU cache hierarchy %
 to absorb and coalesce 
transactional 
updates
and mainly commits them on-chip.
It gains superior efficacy in handling continuous transactions with such a spacious transient persistence domain.

{The throughputs of Kiln and TC are limited by two factors. 
One is due to STT-RAM's longer write/read latencies.  
The other one 
is that  
Kiln and TC have enforced limits in using persistent caches,
such as using a small side-path cache~\cite{txm:transaction-cache:DAC17} or 
 taking a fall-back path to write pmem for backup in case of 
 an almost full request queue~\cite{txm:Kiln:MICRO13}.}
{Other hardware designs,
such as HOOP, Proteus, and FWB, 
exploit hardware components like undo/redo log buffers, WPQ, or  
pmem to  
compose and persist backup copies.  
Continuous transactions keep  limited log buffers or WPQ entries 
being fully occupied over time. 
They are hence inferior to \Attend that leverages the extensive
CPU cache hierarchy to absorb and process data.}

{\autoref{fig:micro-txThroughput} exhibits different observations across benchmarks, as 
their transactions are  with different semantics and complexities.
For example, 
two insertions with
Linked List and RB-Tree 
differ a lot.
Unlike prior designs varying significantly across benchmarks,  
\Attend shows consistently superior performance.
Its
strong robustness 
is mainly accredited to  
CPU cache  it exploits.
A multi-level cache hierarchy has
 been proved to be effectual  
for various workloads over decades.}

{\bf Pmem writes.}\hspace{0.5ex}
{To minimize pmem writes
is another  
goal of \Attend.
\autoref{fig:micro-txwrites} captures the quantity of pmem writes caused
in running transactions with micro-benchmarks (Y axis in the logarithmic scale).
\Attend significantly reduces pmem writes. On average,
the data it writes is   
29.8\%, 37.4\%, 5.3\%, 1.4\%, 1.5\%, 
and 2.1\% that
of other designs in the foregoing order, respectively.
Write-backs of \Attend only happen upon normal cache replacement or
power-off, so it performs pmem writes in a passive and lazy way. 
SWL\_eADR incurs 
double writes.
Kiln takes a  
fall-back path that forcefully  
sends 
 cache lines to pmem with an overflowing request queue.
TC has a similar fall-back path to write pmem
when its side-path cache is almost full. Also, 
whenever a transaction commits, TC issues relevant stores to pmem.
Other hardware designs 
explicitly write backup copies
to pmem.  
 HOOP,
for example, does address indirection at the MC  
and migrates data between pmem locations over time
for out-of-place updating. Proteus leverages the
WPQ for buffering to reduce pmem writes, but the limited capacity of WPQ 
impedes its efficacy.
By using CPU cache in numerous megabytes
to absorb and coalesce data updates,  \Attend 
effectively
minimizes pmem writes.}

{{\bf Tail latencies.} 
Without loss of generality,	
	we record  
99P/99.9P (99- and 99.9-percentile) 
 tail latencies for transactions and
 show them for three benchmarks
in \autoref{fig:99p} and \ref{fig:999p}, respectively, with
Y axes in the logarithmic scales.
These results further justify \Attend' efficacy 
as it  makes much shorter tail latencies  
with various workloads. Take B+-Tree for example.
\Attend' 99.9P tail latency is 40.5\% and 55.1\%
shorter   
 than that of Kiln 
and HOOP, respectively.}

\begin{figure}[b]
	\vspace*{-3ex}
	\centering
	\begin{subfigure}[t]{0.49\columnwidth}
		\centering
		\includegraphics[width=\columnwidth]{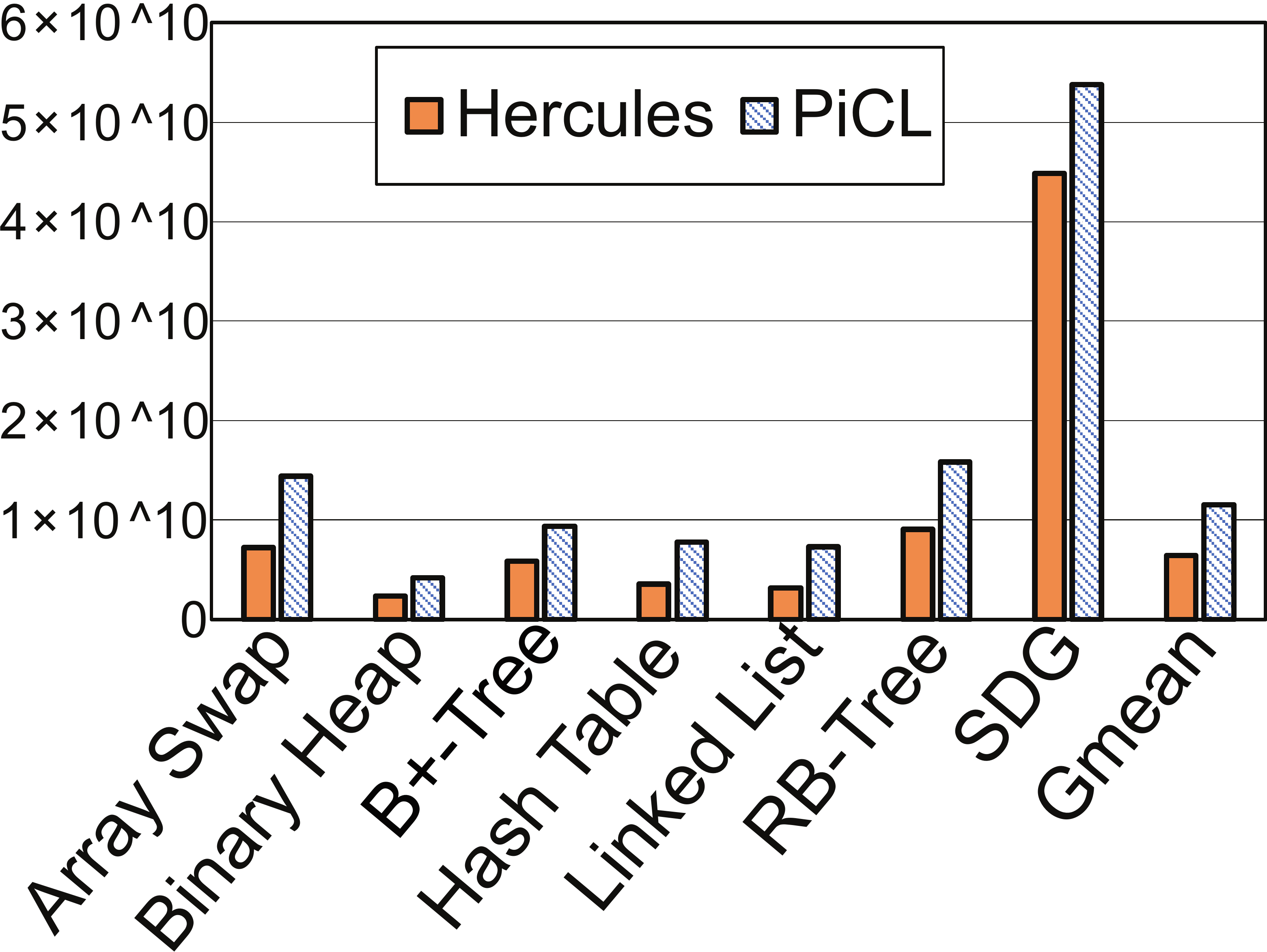}
		\caption{CPU Clock Cycles}
		\label{fig:picl-cycles}
	\end{subfigure}
	\hfill
	\begin{subfigure}[t]{0.49\columnwidth}
		\centering
		\includegraphics[width=\columnwidth]{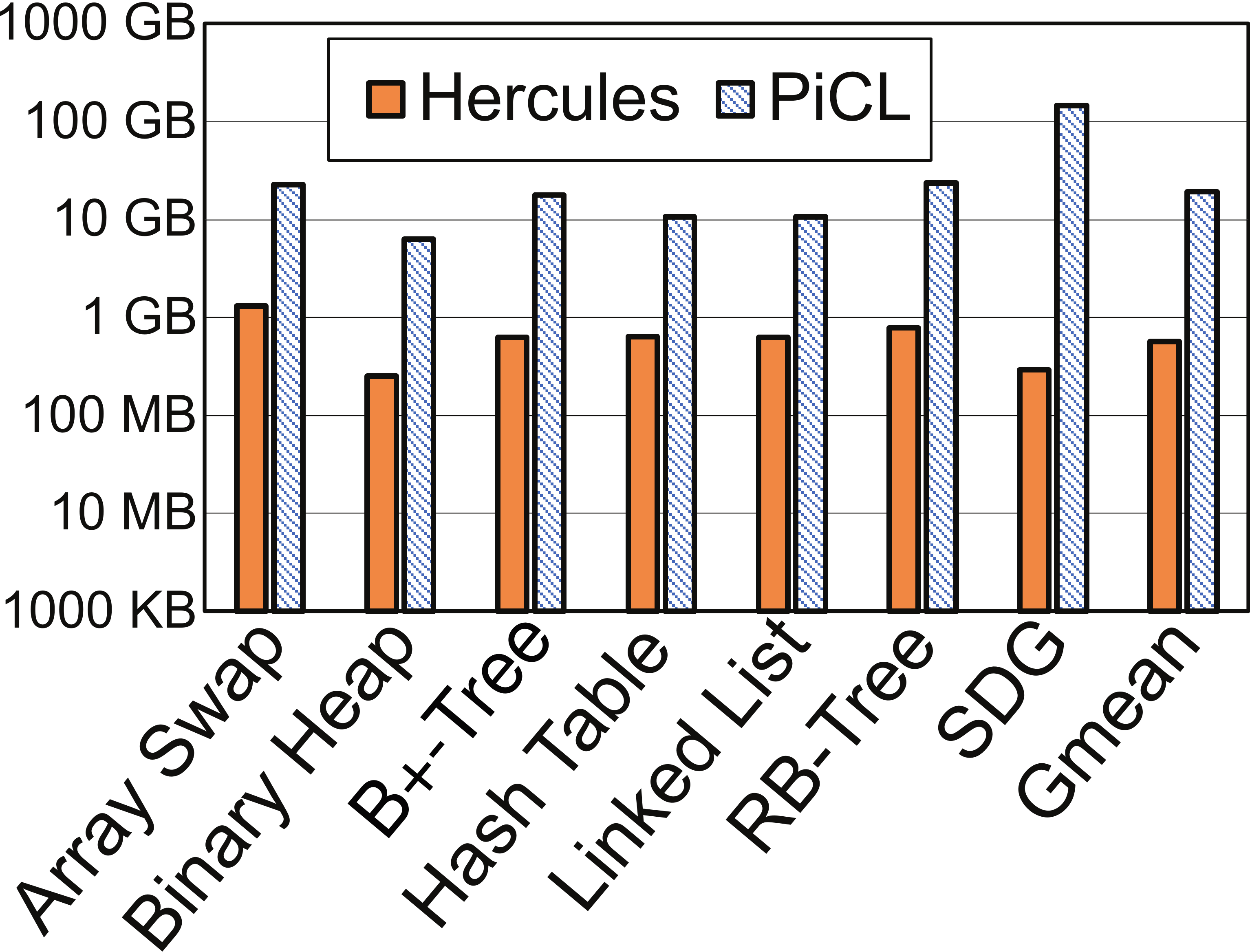}
		\caption{Pmem Writes}
		\label{fig:picl-writes}
	\end{subfigure}
	\vspace{-1ex}
	\caption{A Comparison between \Attend and PiCL}\label{fig:eval-PiCL}
\end{figure}

{\bf A comparison to the approach of checkpointing.}\hspace{0.5ex}
PiCL checkpoints  
data periodically
in pmem 
for 
recovery without forming  
transactions~\cite{txm:checkpoint-SST-RAM:ICCAD-2014,txm:PiCL:MICRO18}.
{In \autoref{fig:eval-PiCL}
 we present the number of 
clock cycles and the quantity of pmem writes
\Attend and PiCL have done to complete all operations for each benchmark.
PiCL's checkpointing is like 
undo-logging
all data at the start of every epoch,   
which, albeit being simplistic, causes  
on average
78.3\% more time 
and 
34.1$\times$ pmem writes  
than \Attend.  
\Attend only covers 
data programmers place
in fine-grained 
transactions and leverages CPU cache to log and buffer them.
This is why
\Attend costs both much less time and  
dramatically fewer pmem writes than PiCL.}

\begin{table}[h]
	\vspace{-1ex}	
	\centering
	\caption{{A Summary of {\tt TxLen}s for Micro-benchmarks}}
	\label{tab:txn-line-num}
	\vspace{-1ex}
	\resizebox{\linewidth}{!}{	
		\begin{tabular}{|l|c|c|c|c|c|c|c|}  
			\hline
			\multirow{2}{*}{Benchmark}&\multicolumn{1}{c|}{Array}    & \multicolumn{1}{c|}{Binary} & B+- & \multirow{1}{*}{Hash}  & \multicolumn{1}{c|}{Linked}  & \multicolumn{1}{c|}{RB-} & \multirow{2}{*}{SDG} \\ 
			 &\multicolumn{1}{c|}{Swap}&\multicolumn{1}{c|}{Heap}&\multicolumn{1}{c|}{Tree}&\multirow{1}{*}{Table} &\multicolumn{1}{c|}{List} &\multicolumn{1}{c|}{Tree}&\\ \hline	
			{Min. {\tt TxLen} } & {7} & {8} & {15} & {3}  & {12}  & {2}  & {22}   \\ \hline
			{Max. {\tt TxLen} } & {9} & {1,021} & {143} & {7}  & {18}  & {71}  & {149}   \\ \hline			
			Avg. {\tt TxLen} &	8.0 & 11.6 & 34.4& 6.2 & 12.5 &50.6 & 33.8 \\ \hline			
		\end{tabular}
	}
	\vspace{-2.5ex}
\end{table}

\begin{figure}[b]
	\vspace*{-3ex}
	\centering
	\begin{subfigure}[t]{0.49\columnwidth}
		\centering
		\includegraphics[width=\columnwidth]{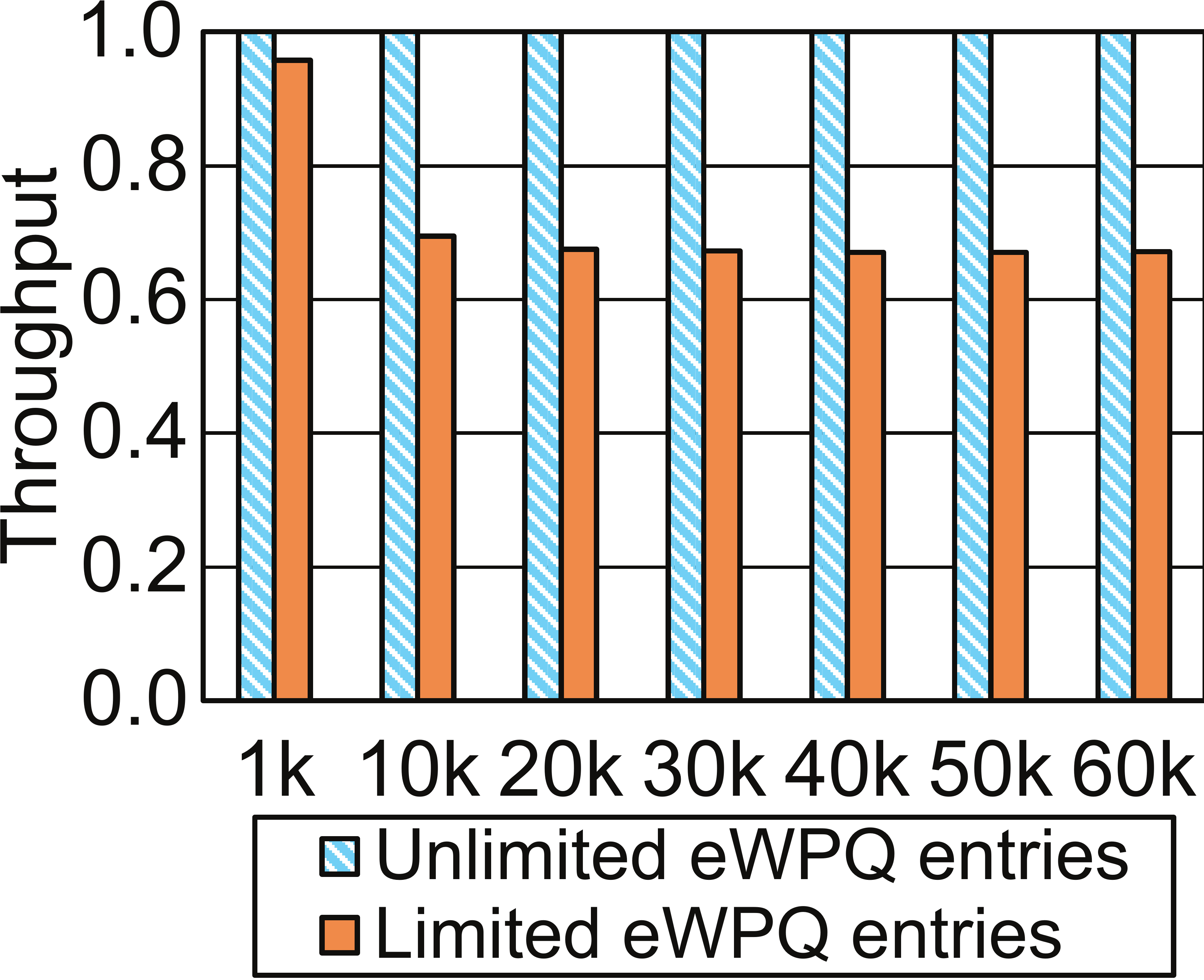}
		\caption{Varying Transaction Sizes}
		\label{fig:txns-large}
	\end{subfigure}
	\hfill
	\begin{subfigure}[t]{0.49\columnwidth}
		\centering
		\includegraphics[width=\columnwidth]{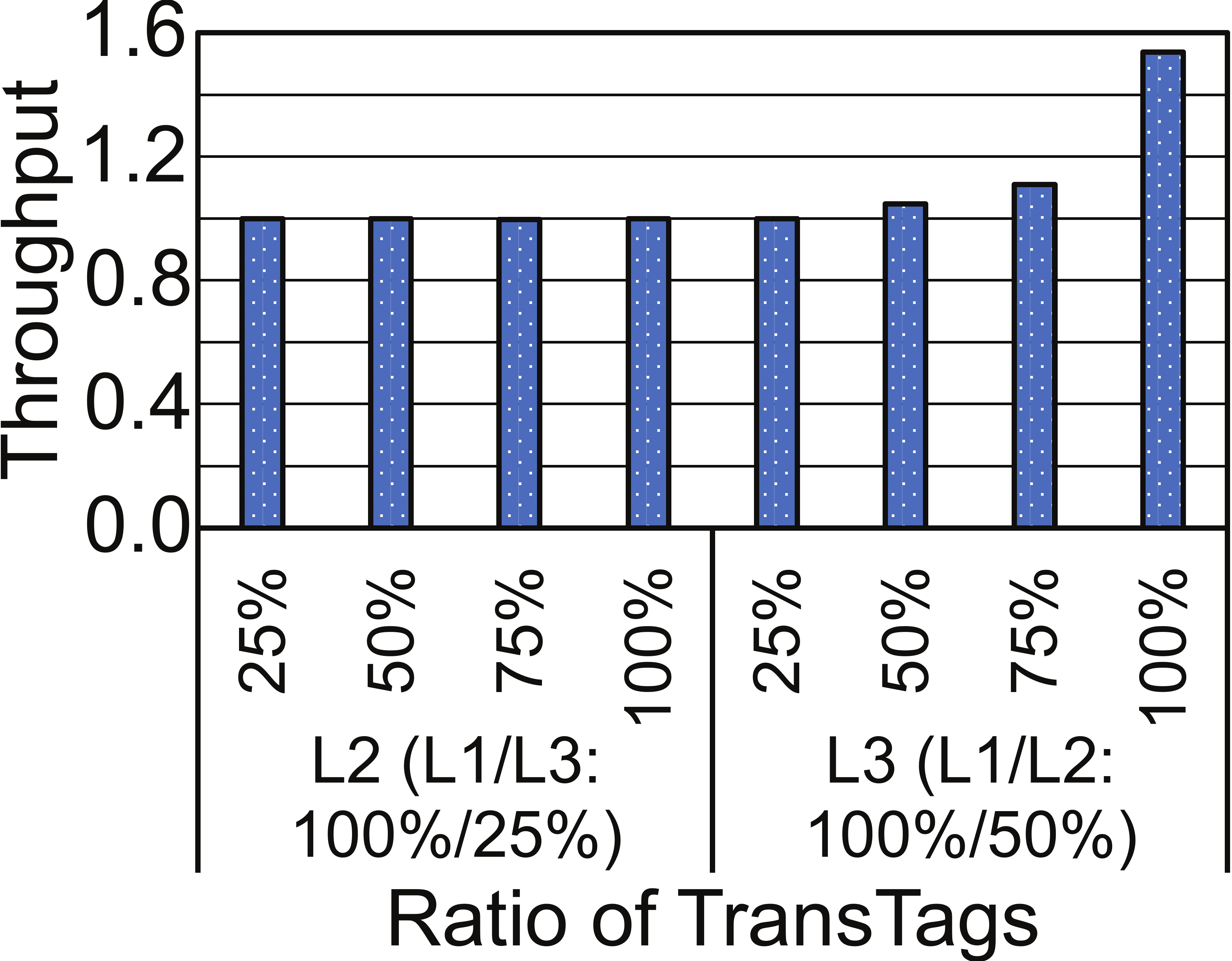}
		\caption{Impact of TransTag Ratios}
		\label{fig:txn-ratio-30k}
	\end{subfigure}
	\vspace{-1ex}
	\caption{The Throughputs with Artificial Huge Transactions}\label{fig:eval-ratio}
\end{figure}

\subsection{Tests for Premature Flush and TransTag Ratios }\label{sec:ratio-eval}
{We summarize {\tt TxLen}s  
on finishing all transactions for each micro-benchmark
in \autoref{tab:txn-line-num}}.
Modern cache hierarchy
has no difficulty in putting  
few to dozens of 
cache lines.
\Attend hence effectually suits
ordinary workloads and  
almost all transactions can be committed on-chip.  
Meanwhile, 
ordinary transactions in small sizes 
are hardly affected by 
 the change of 
TransTag ratios and seldom cause premature flushes or GC.

{To thoroughly evaluate \Attend, we run
unrealistic artificial tests executed in a shrunk 
cache hierarchy.
We scale down  
L1D/L2/L3
caches by 8$\times$ 
to be
3840B/32KB/2MB and synthesize  
test cases in which a huge transaction  
updates
massive  
cache lines.}  
{We
set the length  
of a transaction to be 
1k, 10k, 20k to 60k (k: $\times10^3$)
and run  
with  
unlimited and 512 eWPQ entries,
respectively.
\autoref{fig:txns-large} shows the throughputs normalized against
that with 
unlimited 
eWPQ.}
{For a huge transaction involving 
tens of thousands of cache lines, CPU cache and eWPQ 
would be
saturated and \Attend must use the in-pmem eWPQ extension.
The impact of premature flushes grows up.  
For example,
the throughput with 512 eWPQ entries
declines
by 32.8\% at the 60k case.}

{We also run unrealistically huge transactions  
when varying TransTag ratios at L1D, L2, and L3.}
Without loss of generality, we illustrate with the 30k test
case upon changing
TransTag ratios at L2 and L3.
In \autoref{fig:txn-ratio-30k}, we 
normalize the throughputs against that of 
  default 100\%/50\%/25\%
ratios. 
There is an evident uptrend along an increasing ratio at L3. 
A 2MB L3 cache has 32,768 cache lines.
An increased ratio means more space to house the working set of 30k case.
Only   100\% ratio at L3 manages to
fit all 30,000 cache lines of 30k
and yields the highest throughput without 
premature flush.

\begin{figure}[t]
	\begin{center}
		\scalebox{.94}{\includegraphics[width=\columnwidth]{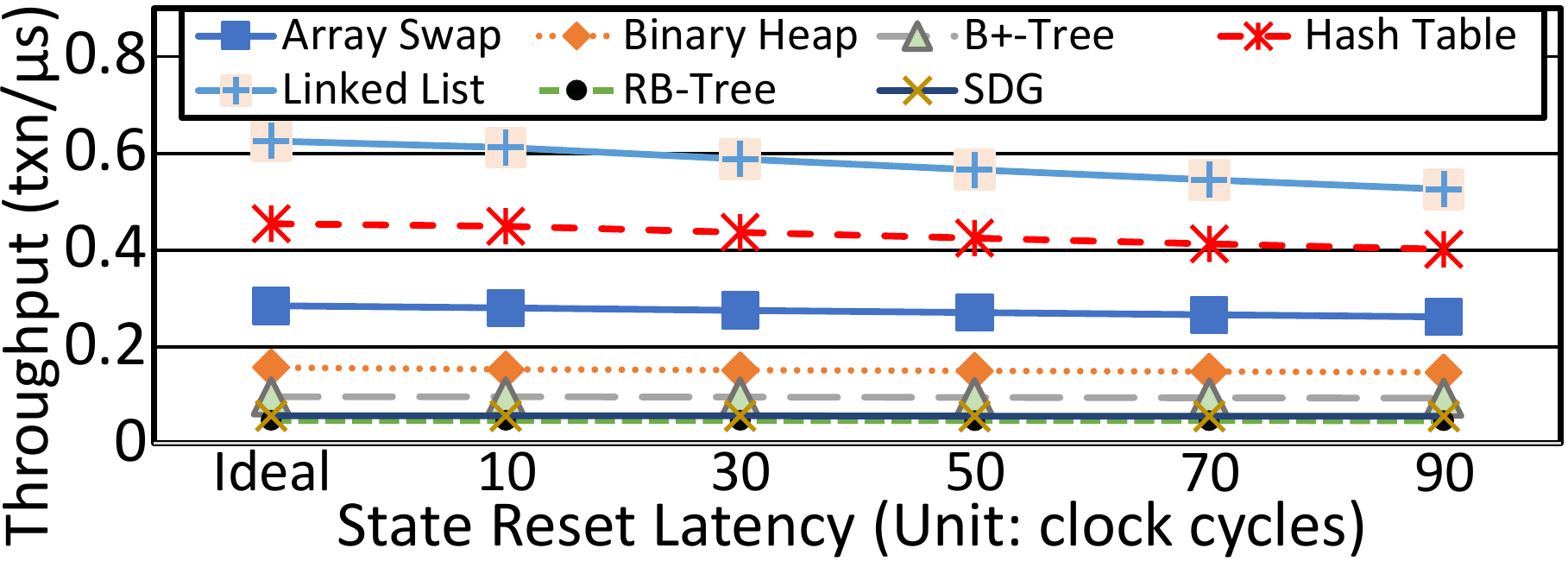}}		
		\vspace{-1ex}
		\caption{{The Impact of} {State Reset Latency}}
		\label{fig:eval-commit-cycle}
	\end{center}
	\vspace{-5ex}	
\end{figure}

\begin{figure*}[t]
	\begin{minipage}[b]{0.51 \columnwidth}
	\includegraphics[width=\textwidth,page=1]{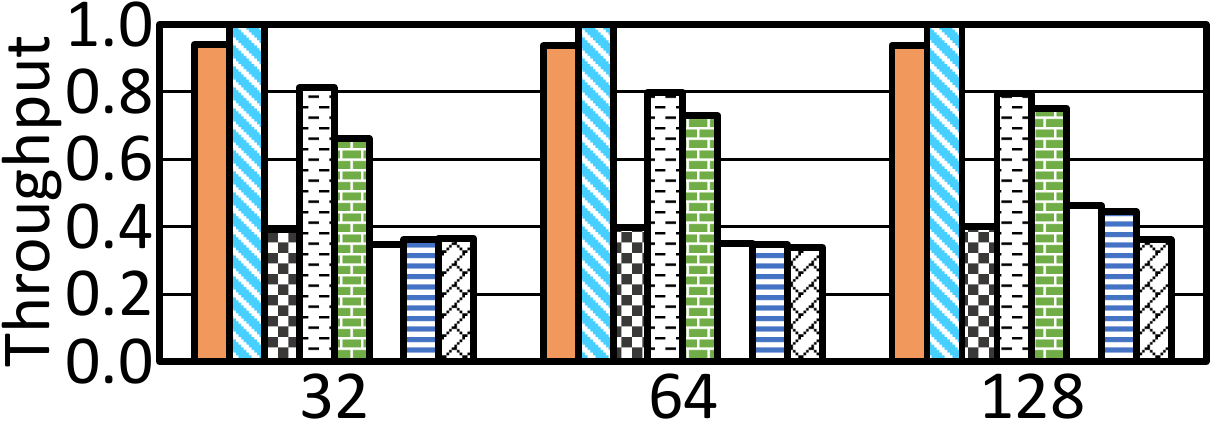}  
	\caption{{The Impact of WPQ Size on Linked List}}\label{fig:wpq-impact} 
	\end{minipage}	
	\hspace{0.01\textwidth}
	\begin{minipage}[b]{1.491\columnwidth}
		\includegraphics[width=\textwidth,page=3]{./micro-macro.pdf} 
		\caption{A Comparison on Throughput of Macro-benchmarks with One Thread (Normalized against OPT's)}		\label{fig:macro-txThroughput}  
	\end{minipage}
	\vspace*{-5ex}
\end{figure*}

\begin{figure*}[t]
	\begin{minipage}[b]{0.51 \columnwidth}
		\includegraphics[width=\textwidth,page=1]{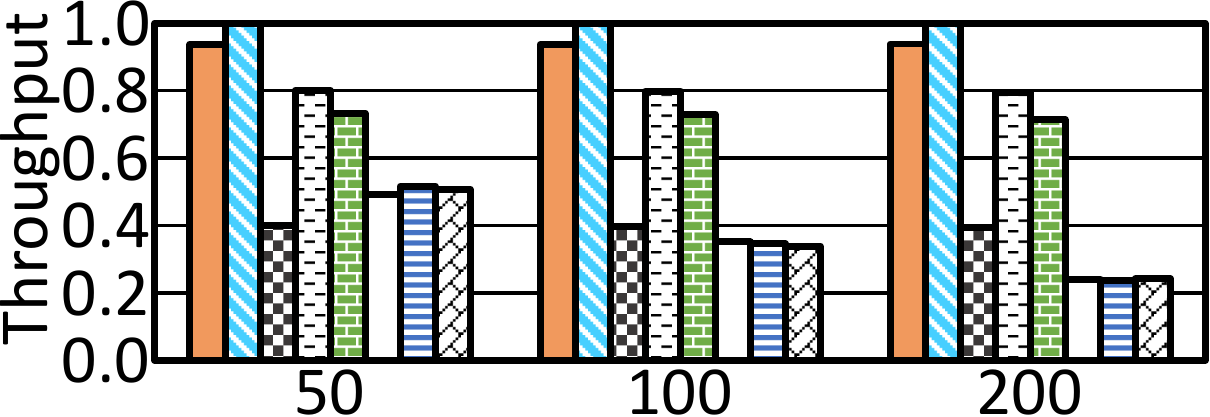}  
		\caption{{The Impact of Pmem Latency (ns) on Linked List}}\label{fig:latency-impact} 
	\end{minipage}
	\hspace{0.01\textwidth}
	\begin{minipage}[b]{1.491\columnwidth}
	\includegraphics[width=\textwidth,page=4]{./micro-macro.pdf} 
	\caption{A Comparison on Throughput of Macro-benchmarks with Four Threads (Normalized against OPT's)}\label{fig:macro-4thread-txThroughput}
\end{minipage}
\vspace*{-6ex}
\end{figure*}

\subsection{{The Impacts of Factors on \Attend}}\label{sec:scan-eval}

{{\bf State reset latency.}} \hspace{0.5ex}
{We set the default state reset latency per transaction
as ten clock cycles} {to toggle {\tt TxState}s of aggregated TransTags
with auxiliary circuit}.
{We can deploy circuits in different complexities
to gain different latencies.}
{\autoref{fig:eval-commit-cycle} shows the throughput (txn/$\mu$s) curves  
on running micro-benchmarks when we vary the latency duration from the ideal (zero)
to 90 cycles.
With a longer 
latency,
the throughputs of Linked list, Hash Table, Array Swap, and Binary Heap 
decrease more severely than those of B+-Tree, RB-Tree, and SDG.
The curves in \autoref{fig:eval-commit-cycle} 
complement {\tt TxLen}s in~\autoref{tab:txn-line-num}.
The throughput of 
a benchmark with
smaller transactions
is   surely
 more sensitive to the increase of state reset latency, since
a smaller transaction itself takes less execution time. 
Take RB-Tree and Linked List
for comparison again. 
An insertion to RB-Tree is generally more complicated and
involves 
50.6 transactional cache lines per transaction.  
Yet an insertion to Linked List deals with  
12.5 cache lines  
per transaction on average.
As a result, increasing the state reset latency more affects lighter
benchmarks like Linked List.}

{{\bf WPQ size.}}  
{We further test when varying the size of WPQ. 
Without loss of generality, 
we choose Linked List
and present 
 throughputs of all designs in~\autoref{fig:wpq-impact}.
\Attend' bottleneck is not on  
write-backs via the WPQ to pmem, 
so its performance has no evident fluctuations.
This justifies the robustness of \Attend in utilizing extensive CPU cache for atomic durability.}
{A	larger WPQ helps prior works like HOOP and Proteus
	yield performance improvements.
Proteus leverages the WPQ of MC for transactional operations while HOOP 
depends on the MC to do address indirection and data movements.
They hence benefit more from an increase  
of WPQ entries.}

{{\bf Pmem latency.}}\hspace{0.5ex}{The default write latency of pmem is 100ns. We vary it 
	to emulate different pmem products.
\autoref{fig:latency-impact} captures
the throughputs of all designs again on Linked List with three write latencies.
By keeping all transactional data in  
CPU caches
without incurring
pmem writes, \Attend is unaffected
by varying pmem latencies. Notably,
the performances of prior works that rely on writing data to
pmem for backup
badly degrade. In summary,
leveraging the capacious transient
persistence domain made of CPU caches
 enables \Attend with high 
adaptability to various pmem products.}

\begin{table}[h]
\vspace{-1ex}
	\centering
	\caption{{The Spatial Cost of \Attend}}\label{tab:spatial_cost}
	\vspace{-1ex}
	\resizebox{\linewidth}{!}{
		\begin{tabular}{|c|c|c|c|c|c|}
			\hline
			\multicolumn{3}{|c|}{TransTags (Unit: KB)} & \multicolumn{3}{c|}{eWPQ (Unit: Bytes)} \\ \hline
			{L1D} & {L2} & {L3} & {Entries} & {Validity Bitmap} & {Pointers} \\ \hline
			{1.29 (per core)} & {6.25 (per core)} & {208} & {4,096} & {64} & {16}  \\ \hline
		\end{tabular}
	}
	\vspace{-3ex}
\end{table}

\subsection{Recoverability, Energy and Spatial Costs of \Attend}\label{sec:energy}

{\bf Recovery.}\hspace{0.5ex}
We tailor gem5's {\em checkpoint} function to save all   
metadata and data into the log zone backed by files  
and emulate a crash 
by encountering  
{a simulator magic
 instruction}.  
 \Attend manages to 
recover properly and resume execution.

{\bf Spatial cost.} 
\autoref{tab:spatial_cost} 
{summarizes the overall
spatial cost of \Attend such that
 removing 2KB at L1D per core
is  
sound to evaluate it}.
	{Similar to on-chip buffers used by prior works~\cite{arch:BBB:HPCA-2021,txm:transaction-cache:DAC17,txm:ReDU:Micro-2018,txm:PiCL:MICRO18,txm:Steal-but-No-Force:HPCA18,txm:MorLog:ISCA-2020},
TransTags incur the main on-chip spatial cost for \Attend.} 
Due to 
the space limitation,
we brief 
 an estimate
with a 
 4-way 
  L1D cache in 30KB.
As we use all L1D cache lines,  
 a TransTag needs 22 bits (21-bit {\tt TxID}
and 1-bit {\tt TxState}) without {\tt WayNo}.
The original metadata per line, 
such as cache tag and state,
takes  
 at most 
 48 bits for a VIPT cache \cite{arch:SIPT:HPCA-2018,arch:SEESAW:ISCA-2018}.
 TransTags thus 
 cost $3.9$\% ($\frac{\textnormal{22}}{48 + 64\times8}$) more
space. 
{This also explains why we increase the   
tag latency by 30\%  
on accessing a transactional cache line ($\frac{\textnormal{22}}{48 + 22}$ $\approx$ 31.4\%).
Similarly we estimate the proportions of {TransTag}s at L2 and L3 to be  
$2.2$\% and $1.2$\%, respectively.}  
 In all,
the spatial cost of \Attend is insignificant.  

{\bf Energy cost.}\hspace{0.5ex}
If a crash occurs,
besides 
the eADR's ordinary flushes,  
\Attend   
persists the TransTag and home address
that are estimated as at 
most \FPeval{\Vcache}{10}\Vcache{}B (e.g., $\ge\frac{22 + 48}{8}$ at L1D)
for a 
transactional cache line.
The energy costs per store 
from L1D, L2, and L3 caches 
 to pmem on a crash 
are respectively $11.839nJ/$B, $11.228nJ/$B, and $11.228nJ/$B \cite{arch:BBB:HPCA-2021, IISWC:energy:energy-cost}.
The base cost of flushing the entire
 cache hierarchy 
is thus
\FPeval{\Vorig}{round((11.839 * 30 * 1024 * 8 + 11.228 * 256 * 1024 * 8 + 11.228 * 16 * 1024 * 1024) / 1000000, 3)}\Vorig{}$mJ$ for 
8-core CPU.
{Regarding}
\FPeval{\VLone}{round(30 * 1024 / 64 * 1 * 8, 0)}\VLone{}/
\FPeval{\VLtwo}{round(256 * 1024 / 64 * 0.5 * 8, 0)}\VLtwo{}/
\FPeval{\VLLC}{round(16 * 1024 * 1024 / 64 * 0.25, 0)}\VLLC{}
{transactional cache lines at L1D/L2/L3 with eight cores},  
all of them
  are dirty and uncommitted in the worst case.
The 
cost to flush TransTags and home addresses 
 for them is about
\FPeval{\Vaddi}{round(((11.839 * \VLone{} + 11.228 * \VLtwo{} + 11.228 * \VLLC{}) * \Vcache{}) / 1000000, 3)}\Vaddi{}$mJ$.
\Attend also needs to  identify transactional cache lines by
comparing {\tt WayNo} in each {TransTag} and 
all comparisons cost about
\FPeval{\Vcomp}{round((\VLone{} + \VLtwo{} + \VLLC{}) * 0.98 / 1000, 3)}\Vcomp{}$nJ$
with an up-to-date comparator taking $0.98 pJ$ per comparison~\cite{energy:Comparator}.
Overall,
\Attend maximally brings about 
 \FPeval{\Vres}{round((\Vaddi{} + \Vcomp{} / 1000000)/ \Vorig{} * 100, 2)}\Vres{}$\%$
($\frac{\Vaddi{}+\Vcomp{} \times 10^{-6}}{\Vorig{}}$)
extra energy cost,  
which
is practicable 
in upgrading the eADR.
In addition, a {\tt TxID} in more bits  
may increase such extra  
cost to be beyond 
5.0\% or even greater,
and also impose 
further challenges on designing and producing chips with
 efficiency and reliability~\cite{arch:Error-control:TCAD-2005,arch:circuit-design:IEEE-2008,arch:reliable:HPCA-2022,arch:NVMExplorer:HPCA-2022}.
 
\subsection{Macro-benchmark}\label{sec:macro-eval}

We utilize TPC-C and YCSB workloads
from WHISPER~\cite{bench:YCSB:SoCC-2010,txm:ReDU:Micro-2018,arch:Dolos-secure:Micro-2021} to evaluate \Attend
for two purposes.
Firstly, we measure the performance and robustness of \Attend
with more sophisticated  
workloads  
of 
realistic applications.
Secondly, we run with multi-threads   
to test the 
scalability of \Attend in serving concurrent transactions.
TPC-C's New-order 
 follows the de facto 
 semantics. 
As to YCSB,
we vary the value size   
for a comprehensive test.

As shown in \autoref{fig:macro-txThroughput},  
on dealing with more complicated  
transactions of macro-benchmarks, 
\Attend still  
yields a higher throughput
than SWL\_eADR, Kiln, TC, HOOP, Proteus, and FWB by 39.2\%, 20.5\%, 19.0\%, 31.5\%,
41.1\%, and 41.7\% on average, respectively. 
 Furthermore, with a larger value, the performance gap between
 \Attend and prior designs generally becomes wider. 
The root cause is \Attend' robust design.
It leverages the sufficient CPU cache
 to  take in transactions.
The eADR renders a value almost persistent in CPU cache.
A larger value does not greatly increase 
the cost of persisting the
value, especially on the critical path.
Comparatively,
the double writes of logging 
again hinder SWL\_eADR from achieving high performance.
As to Kiln and TC, a larger value imposes more burdens in using FIFO 
queues and side-path cache, respectively,
 and triggers more executions through  
 fall-back paths
that severely affect their throughputs.
For FWB and Proteus, 
larger  
transactions still make immense
data updates that 
continually run out of 
their log buffer and WPQ entries, respectively.   
As to HOOP, larger values consume more pmem bandwidths and incur longer time for the MC to wait for the completion of data migrations.

The results with running four threads
 in \autoref{fig:macro-4thread-txThroughput} 
 justify the scalability of \Attend, which produces 43.7\%, 
20.1\%, 27.8\%, 31.6\%, 42.9\%, and 43.7\%
higher throughput, respectively, than prior  
designs 
on average. 
{A multi-level private/shared
	 cache hierarchy with the set-associativity management 
implies an innate scalability to support multi-threading
with multi-cores.} 
\Attend gains scalability accordingly.
Threads may share {\tt GlobalTxID} register at the start of a transaction and
 eWPQ outside of the critical path 
 for prematurely flushed 
cache lines. Though, 
getting a {\tt TxID} can be swiftly done 
 in an atomic operation
 while 
the  probability 
of massive synchronizations 
under a spacious 
  cache hierarchy is low.
For prior designs, the contention on resources
between multi-threads is much fiercer.
 Take HOOP for illustration 
  again. 
 As 
 it depends on the MC for  
 out-of-place
 data updates,  
multiple threads contend flushing data through
the MC.   
This offsets the effect of CPU cache for concurrency
  and makes the MC 
   a busy synchronization point being shared at runtime,
   thereby limiting HOOP's scalability.

\section{Conclusion}\label{sec:conclusion}

We propose \Attend,
a systematic hardware
 design 
leveraging the transient persistence domain made of
CPU cache 
to enable  
the transaction-level atomic durability for 
 in-pmem data. 
\Attend has comprehensive control logics and data-paths installed 
 in CPU cache, MC, and pmem. It provides
transactional primitives and protocols %
to define and proceed
transactions.
\Attend well serves typical applications.
Experiments confirm that
it significantly outperforms prior works  
with higher performance. 
\Attend also substantially 
minimizes pmem writes
with ample CPU cache   
buffering data.   

\balance
\bibliographystyle{IEEEtranS}
\bibliography{arch}

\end{document}